\documentclass[aps,prd,twocolumn,10pt,superscriptaddress]{revtex4-1}
\pdfoutput=1
\usepackage{graphicx}
\usepackage{booktabs}
\usepackage{color}
\usepackage[breaklinks,colorlinks]{hyperref}
\hypersetup{
    linkcolor={black},
    citecolor={black},
    urlcolor={black}
}
\usepackage{amssymb,amsmath,amsfonts}
\usepackage[utf8]{inputenc}

\begin{document}


\title{Baryon number fluctuations in  chiral effective models and their phenomenological implications  }

\author{G\'abor Andr\'as Alm\'asi}
\email[]{g.almasi@gsi.de}
\affiliation{GSI Helmholtzzentrum f\"{u}r Schwerionenforschung, D-64291 Darmstadt, Germany}
\author{Bengt Friman}
\affiliation{GSI Helmholtzzentrum f\"{u}r Schwerionenforschung, D-64291 Darmstadt, Germany}
\author{Krzysztof  Redlich}
\affiliation{University of Wroc\l aw, Institute of Theoretical Physics,   PL 50-204 Wroc\l aw, Poland}
\affiliation{ExtreMe Matter Institute EMMI, GSI, D-64291 Darmstadt, Germany}

\begin{abstract}
We study  the critical properties of net-baryon-number fluctuations at the   chiral restoration transition
in  a medium at finite temperature and net  baryon density.
The chiral dynamics of quantum chromodynamics (QCD) is modeled by the Polykov-loop extended  Quark-Meson  Lagrangian,  that includes the coupling of quarks to vector meson and temporal gauge fields. The Functional Renormalization Group is employed to properly account for the  $O(4)$ criticality at the phase boundary. We focus on the  properties and systematics  of ratios of the net-baryon-number cumulants $\chi_B^n$, for ${1\leq n\leq 6}$,  near the phase boundary.
The results are presented in the context of the recent experimental data of the STAR Collaboration on  fluctuations of the net proton number in heavy ion collisions at RHIC. We show that the model results for the energy dependence of  the  cumulant ratios are in good overall agreement with the data, with one exception. At center-of-mass energies below ${19.6\;\mathrm{GeV}}$, we find that the measured fourth-order cumulant deviates considerably from the model results, which incorporate the expected $O(4)$ and $Z(2)$ criticality. We assess  the influence  of model assumptions and in particular of repulsive vector-interactions, which are used to modify the location of the critical endpoint in the model, on the cumulants ratios.
Finally, we discuss a possibility to test to what extent the fluctuations are affected by nonequilibrium dynamics by comparing certain ratios of cumulants.

\end{abstract}


\maketitle

\hypersetup{
    linkcolor={blue},
    citecolor={blue},
    urlcolor={blue}
}

\section{Introduction \label{sec:Introduction}}

It is well established  within  lattice QCD (LQCD), that at  vanishing  and small  chemical potential,  $\mu_B$,  strongly interacting matter undergoes a smooth  crossover
from the  hadronic phase to a quark-gluon plasma~\cite{Aoki:2006we,Aoki:2009sc,Borsanyi:2010bp,Bazavov:2011nk}. Moreover,
arguments are given,  that in the limit of massless $u$ and $d$  quarks,  the crossover transition becomes a genuine second-order chiral phase transition, belonging to the $O(4)$ universality class \cite{Pisarski:1983ms,Ejiri:2009ac}.   The nature of this transition  at higher  net baryon  densities is still not settled by first principle LQCD studies, owing to the sign problem. However, in  effective models of QCD, it is found  that at sufficiently large $\mu_B$,  the system can exhibit a first order  chiral phase  transition. The endpoint of this conjectured transition line  in the  $(T,\mu_B)$- plane, is the chiral \textit{critical endpoint} (CEP)~\cite{Asakawa:1989bq,Barducci:1989wi,Wilczek:1992sf,Halasz:1998qr,Ding:2015ona}.  At the CEP, the system exhibits the 2$^{nd}$ order phase transition, which belongs to the $Z(2)$ universality class \cite{Schaefer:2006ds}.

Various methods have been developed to circumvent the sign problem and perform LQCD calculations at nonvanishing $\mu_B$  \cite{Fodor:2001au,Allton:2002zi,Sexty:2013ica}. However, so far these methods are
restricted to rather small values of  chemical potential, ${\mu_B \ll 3 T_c}$, where ${T_c=\; 154(9) \mathrm{MeV}}$  is   the chiral crossover  temperature at vanishing net baryon density \cite{Borsanyi:2010bp,Bazavov:2011nk}.

Due to the restriction  of present  LQCD calculations to small net baryon densities, effective models  that belong to the same universality class as QCD, e.g., the Polyakov-loop extended Nambu-Jona-Lasinio~\cite{Nambu:1961tp,Nambu:1961fr,Fukushima:2003fw,Ratti:2005jh,Sasaki:2006ww,Roessner:2006xn,Fukushima:2008wg,Sasaki:2007db,Sasaki:2006ws} and Quark-Meson (PQM)  models~\cite{Schaefer:2007pw,Schaefer:2009ui,Mizher:2010zb,Herbst:2010rf,Skokov:2010wb,Skokov:2010sf,Skokov:2010uh,Mitter:2013fxa,Herbst:2013ufa}, have been  employed to study the chiral  phase transition
for a broad range of thermal parameters. Moreover, there are ongoing studies using functional methods that focus on systematic improvements  of such  effective models,  to achieve a more quantitative matching with LQCD, and eventually to perform reliable calcuations of the properties of QCD matter at nonzero net baryon density~\cite{Mitter:2014wpa,Braun:2014ata,Fischer:2014ata}.

One of the strategic goals of current experimental and theoretical studies of chiral symmetry restoration in QCD is to unravel the phase diagram of strongly interacting matter and to clarify whether a chiral CEP exists. A dedicated research program  at RHIC, the beam energy scan (BES), has been established to explore these issues in collisions of heavy ions at relativistic energies \cite{Aggarwal:2010cw}. By varying the beam energy  at RHIC, the properties of strongly interacting matter in a broad range of net baryon densities, correponding to a wide range in baryon chemical potential, ${20\; \mathrm{MeV} <\mu_B<500\; \mathrm{MeV}}$~\cite{BraunMunzinger:2003zd,Andronic:2005yp,Andronic:2008gu,Cleymans:2005xv}, can be studied.
In particular, fluctuations of conserved  charges are considered as relevant probes of the phase diagram~\cite{Stephanov:1998dy,Stephanov:1999zu,Asakawa:2000wh,Jeon:2000wg,Friman:2011pf,Ejiri:2004bh,Allton:2005gk,Ejiri:2005wq,Karsch:2005ps,Karsch:2010ck,Friman:2011pf,Sasaki:2007db,Sasaki:2006ws}.
These  are experimentally accessible and reflect the criticality of the chiral transition. Fluctuations of the net baryon number are particulary interesting, owing to a direct connection to critical scaling near the chiral phase boundary~\cite{Friman:2011pf,Karsch:2010ck}.

First data on net-proton-number fluctuations, which are used as a proxy for fluctuations of the  net baryon number,  in heavy-ion collisions have been obtained by the STAR Collaboration at  RHIC energies~\cite{Adamczyk:2013dal,Luo:2015ewa,Luo:2015doi}. The STAR  data  on the  variance, skewness and kurtosis of net proton number,  are intriguing and have stimulated lively discussions on their physics origin and interpretation. {Recently,  first results on the mean and variance of the net-proton distribution were obtained in heavy ion collisions at the LHC energy  by the ALICE Collaboration}~\cite{Anartalk}.

In the following,
we focus on  the properties and systematics  of the  cumulants of net-baryon-number  fluctuations near the chiral phase boundary. {Our studies are performed  based on}
 the Polykov-loop extended  Quark-Meson  Lagrangian which includes  couplings of quarks to  vector and temporal gauge fields. To account for the  ${O(4)}$ and ${Z(2)}$ critical fluctuations
near the phase boundary, we employ the Functional Renormalisation Group (FRG)~\cite{Wetterich:1992yh,Morris:1993qb,Berges:2000ew,Polonyi:2001se}.
We present contour plots of ratios of cumulants, involving the mean, the variance, skewness and kurtosis in the ${(T,\mu_B)}$- plane. These ratios obtained on the phase boundary and  on a freeze-out line determined by fitting the skewness ratio, following Ref. \cite{Bazavov:2012vg}, are confronted with the corresponding experimental values of the STAR Collaboration.
We explore the  influence of the CEP and the repulsive interactions on the fluctuation observables. The calculations are performed with two different initial conditions and two Polyakov-loop potentials, to assess the model dependence of the results.

We find that the energy dependence of  ratios of   low-order cumulants,  $\chi_B^n$ with ${n=1,2,3}$,
are in good agreement with the data, whereas for the ratio involving the kurtosis, the model results differ from the data at energies below $\sqrt{s}\simeq 20$ GeV, corresponding roughly to the top SPS energy. At these low energies, the latter ratio increases strongly, while the model results, which embody $O(4)$ as well as $Z(2)$ criticality, differ substantially from the data. Finally, we discuss possible caveats, which could undermine our conclusions and assess the uncertainties of the model.

The paper is organized as follows. In Sec.~\ref{sec:PQM} we formulate the model and the FRG method, which is employed to compute  cumulants of net baryon number. In Sec.~\ref{sec:phaseboundary} we present results on the characteristics of susceptibility ratios as functions of temperature and chemical potential and study, along the phase boundary, the dependence of these ratios on the vector interaction. Moreover, in this section we confront the model results with the STAR data on net-proton-number fluctuations. Sec.~\ref{sec:summary} contains  summary and conclusions.

\section{The Polyakov--quark-meson model \label{sec:PQM}}
The PQM model is a low energy effective approximation  to QCD  formulated in terms of the light quark $q=(u,d)$ as well as scalar and the pseudoscalar meson $\phi=(\sigma,\vec{\pi})$ fields. The quarks are coupled to  the  background Euclidean gluon field $A_{\mu}$, with vanishing  spatial components,  which is linked  to  the Polyakov loop
\begin{align}
	\Phi &=            \frac{1}{N_c}\left\langle \mathrm{Tr}_c \mathcal{P} \exp\left( i\int_0^{\beta} d\tau A_0 \right) \right\rangle, \\
	\bar{\Phi} &= \frac{1}{N_c}\left\langle \mathrm{Tr}_c \mathcal{P} \exp\left( -i\int_0^{\beta} d\tau A_0 \right) \right\rangle .
\end{align}
Moreover, we include a coupling  of the quarks to a massive vector field $\omega$.
The resulting Lagrangian of the model reads
\begin{align}\label{lagrangian}
	\mathcal{L} = &\bar{q}\left( i\gamma^{\mu} D_{\mu} -g(\sigma +i\gamma_5 \vec{\tau}\vec{\pi})-g_{\omega}\gamma^{\mu}\omega_{\mu}\right)q
	 \nonumber\\
	&+\frac12 \left(\partial_{\mu}\sigma\right)^2+\frac12 \left(\partial_{\mu}\vec{\pi}\right)^2-U_m(\sigma,\vec{\pi})
	\nonumber\\
	&-\mathcal{U}(\Phi,\bar{\Phi};T)-\frac12 m_{\omega}^2\omega^2+\frac{1}{4}F_{\mu\nu}F^{\mu\nu},
\end{align}
where $D_{\mu}=\partial_{\mu}-iA_{\mu}$,  and $F^{\mu\nu}=\partial^\mu\omega^\nu-\partial^\nu\omega^\mu$. The parameters of the mesonic potential
\begin{equation}
	U_m(\sigma,\vec{\pi}) = \frac{\lambda}{4}(\sigma^2+ \vec{\pi}^2)^2 +\frac{m^2}{2}(\sigma^2+ \vec{\pi}^2) - H\sigma,
\end{equation}
are tuned to vacuum properties of the $\sigma$ and $\vec{\pi}$ mesons (see Appendix~\ref{sec:initc}).
We use the Polyakov-loop potential $\mathcal{U}(\Phi,\bar{\Phi};T)$ determined in \cite{Lo:2013hla}, by fitting quenched lattice QCD results for the equation of state and the Polyakov-loop susceptibilities. The parametrization of  ${\mathcal{U}(\Phi,\bar{\Phi};T)}$ is given in Appendix~\ref{sec:initc}.

We compute the thermodynamic properties of this model including fluctuations of the scalar and pseudoscalar meson fields within the framework of the FRG method. The $\omega$ meson, on the other hand, is treated on the  mean-field level.  The MF treatment of the vector field is justified by recent FRG  results obtained at vanishing chemical potential indicating that the vector meson mass remains above the cutoff during the FRG flow  and that the temperature dependence of the screening mass  is very weak~\cite{Rennecke:2015eba,Eser:2015pka}. Therefore, it is expected that fluctuations of the vector fields decouple from the flow. The Polyakov loop is also treated on the mean-field level. The Polyakov-loop variables and the vector field are tuned such that at the end of the calculation a stationary point of the thermodynamic potential is reached.

In the FRG framework, the effective average action $\Gamma_k$, which  interpolates between the classical and the full quantum action, is obtained by solving the renomalization group flow equation \cite{Wetterich:1992yh}
\begin{equation}\label{eq:FRG-flow}
	\partial_k \Gamma_k[\phi]=\frac12 \mathrm{STr}\left[\left( \Gamma_k^{(2)}[\phi]+R_k \right)^{-1} \partial_k R_k\right],
\end{equation}
where $\phi$ denotes the quantum fields considered, $\mathrm{STr}$ is a trace over the fields, over momentum  and over all internal indices. It also adds fermion contribution to the boson contribution with correct sign and prefactor. The regulator function $R_k$ suppresses fluctuations at momenta below $k$. Thus, effects of fluctuations of quantum fields are included shell by  shell in  momentum space, starting from a UV cutoff scale $\Lambda$. We employ the  optimized regulator introduced by Litim \cite{Litim:2001up,Litim:2000ci}, which yields an algebraic expression for the right hand side of the flow equation, Eq.~\eqref{eq:FRG-flow}.

\subsection{Thermodynamics  at vanishing vector coupling}

For vanishing vector coupling,  the vector fields obviously decouple completely. Hence, in such models the thermodynamics is controlled  by the quark, as well as, the scalar and pseudoscalar fields \cite{Herbst:2010rf,Skokov:2010wb,Herbst:2013ufa,Kamikado:2012bt,Tripolt:2013jra}.
\par\noindent
Using $\left. \Omega_k= \Gamma_k/V\right|_{H=0}$ we obtain  the thermodynamic potential density
\begin{equation}\label{potential}
	\Omega(T,\mu_q) = -P(T,\mu_q) = \Omega_{k=0}(T,\mu_q,\sigma^*,\Phi^*,\bar{\Phi}^*)-H\sigma^* ,
\end{equation}
where $\mu_q=\mu_B/3$ denotes the quark chemical potential, the asterisk denotes the expectation value of the condensates and $\Omega_{k=0}$ is  obtained by solving the flow equation,
\begin{align} \label{eq:floweq}
	\partial_k&\Omega_k(T,\mu_q;\sigma,\Phi,\bar{\Phi}) \nonumber \\
	= &\frac{k^4}{12\pi^2}
	\left\lbrace \frac{3}{E_{\pi}}\coth\left(\frac{E_{\pi}}{2T}\right)+\frac{1}{E_{\sigma}}\coth\left(\frac{E_{\sigma}}{2T}\right)\right.
	\nonumber \\ &-\frac{4 N_c N_f}{E_q}\big( 1-N(T,\mu_q;\sigma, \Phi,\bar{\Phi})-N(T,-\mu_q;\sigma,\bar{\Phi},\Phi)\big)\bigg\rbrace ,
\end{align}
 with  $E_q=\sqrt{k^2+g_s^2\sigma^2}$, $E_{\sigma}=\sqrt{k^2+\Omega_k''}$ and $E_{\pi}=\sqrt{k^2+\Omega_k'/\sigma}$.  Here the prime denotes a derivative with respect to $\sigma$ field, $g_s$ is the Yukawa coupling, while  $N_c=3$ is  the number of colors and $N_f=2$  the number of flavors. The Polyakov-loop modified quark occupation numbers are given by
\begin{align}
	N&(T,\mu_q;\sigma, \Phi,\bar{\Phi}) \nonumber \\&= \frac{ 1+2\bar{\Phi} e^{(E_q-\mu_q)/T} + \Phi e^{2(E_q-\mu_q)/T}}
	{1+3\bar{\Phi} e^{(E_q-\mu_q)/T}+3\Phi e^{2(E_q-\mu_q)/T}+e^{3(E_q-\mu_q)/T}}.
\end{align}

The  flow equation (\ref{eq:floweq}) is solved numerically starting from the initial conditions specified at the momentum scale $k=\Lambda$
\begin{equation}
	\Omega_{k=\Lambda} = \left.U_m(\sigma,0)\right|_{H=0} + \mathcal{U}(\Phi,\bar{\Phi};T).
\end{equation}
The thermodynamic potential $\Omega(T,\mu)$, shown in Eq.~\eqref{potential}, is obtained by evaluating the fully evolved $\Omega_{k=0}$ at the stationary point,  $SP=(\sigma^*, \Phi^*,\bar{\Phi}^*)$, determined by
\begin{equation}
	\left.\frac{\partial \Omega_{k=0}}{\partial \sigma}\right|_{SP}= H,\quad
	\left.\frac{\partial \Omega_{k=0}}{\partial \Phi}\right|_{SP}=0, \quad
	\left.\frac{\partial \Omega_{k=0}}{\partial \bar{\Phi}}\right|_{SP}=0.
\end{equation}

Any thermodynamic quantity can be obtained  by taking the appropriate derivatives of the thermodynamic potential~\eqref{potential}. However, quantities that require  higher order derivatives are difficult to extract by numerical differentiation, owing to limited numerical precision. We find that improved numerical stability is obtained when the derivatives up to second order,
\begin{align} \label{eq:flows}
	&\frac{\partial \Omega_k}{\partial \Phi},&\quad &\frac{\partial \Omega_k}{\partial \bar{\Phi}},&\quad
	&\frac{\partial \Omega_k^2}{\partial \Phi^2},&\quad &\frac{\partial \Omega_k^2}{\partial \bar{\Phi}^2}, \quad
	\frac{\partial \Omega_k^2}{\partial \Phi \partial \bar{\Phi}},\nonumber\\
	&\frac{\partial\Omega_k^2}{\partial \mu \partial \Phi},&\quad &\frac{\partial\Omega_k^2}{\partial \mu \partial \bar{\Phi}},&\quad
	&\frac{\partial \Omega_k}{\partial \mu},&\quad
	&\frac{\partial^2 \Omega_k}{\partial \mu^2},
\end{align}
are computed directly using flow equations, summarized  in Appendix~\ref{sec:flows}.

We solve  the coupled flow equations numerically by discretizing them on a grid of the order parameter $\sigma$. In order to recover the Stefan-Boltzmann limit, it is necessary to include the effect of fermion fluctuations above the UV cutoff  $\Lambda$. This is done by adding the contribution of the thermal Polyakov-loop suppressed massless quark loop, integrated from the UV cutoff to infinity~\cite{Skokov:2010wb}.
The initial conditions for the flow equations are fixed by vacuum criteria, as discussed in Appendix~\ref{sec:initc}.

It should be noted that we also performed the RG calculations by expanding the potential in terms of Chebyshev polynomials~\cite{Boyd}, instead of working on a uniformly discretized grid. This method is illustrated in Refs.~\cite{Borchardt:2015rxa,Borchardt:2016pif,Almasi:2016zqf}. Using the same initial conditions it yields the same results for all thermodynamic quantities, but up to a higher numerical precision. This method was required to extract cumulants of high order.

\subsection{Thermodynamics at  nonvanishing vector coupling}
For  nonvanishing vector coupling, the $\omega$ field is coupled to the chiral sector through the fermionic part  of the flow. The temporal component of the $\omega$ field gains a nonvanishing  expectation value at nonzero  density, while the spatial components vanish in a system with zero net quark current.

An inspection of the  Lagrangian \eqref{lagrangian}, reveals that a nonvanishing $\omega$ field effectively acts as a shift of the quark chemical potential. Thus, it is convenient to introduce an effective quark chemical potential, $\nu = \mu -g_{\omega} \omega$~\footnote{In the following,   $\omega$ represents the expectation value of the Euclidean zero-component field, which differs from the Minkowski value by a factor of $i$},
and continue using the formalism for vanishing vector coupling, albeit with modified initial conditions for the flow equation
\begin{equation}
	\Omega_{k=\Lambda} = \left.U_m(\sigma,0)\right|_{H=0} + \mathcal{U}(\Phi,\bar{\Phi};T)-\frac12 m_{\omega}^2 \omega^2.
\end{equation}
Since, for a given value of $\omega$, this amounts to constant shift of the thermodynamic potential, it will not modify the RG flow. Thus, the only effect of a nonzero $\omega$ field on the flow is the replacement of the chemical potential~$\mu$ by the previously defined effective chemical potential~$\nu$.

The expectation value of  $\omega$  is obtained by extremizing $\Omega_{k=0}$. Since the flow for  vanishing  and nonvanishing vector coupling can be related to each other, one can  express all results in terms of quantities computed at vanishing  vector coupling. We denote the thermodynamic potential density and net baryon density at vanishing vector coupling by $\tilde{\Omega}(T,\nu)$ and $\tilde{n}(T,\nu)$,
respectively.  The  expectation value of the $\omega$ field is then determined by
\begin{equation}
	\left.\frac{\partial \Omega_{k=0}}{\partial \omega}\right|_{SP}=0,\quad \Longrightarrow\quad g_{\omega} \omega = -G_\omega \frac{\partial \tilde{\Omega}}{\partial \nu}= G_\omega \tilde{n}(T,\nu),
\end{equation}
where $G_\omega = g_{\omega}^2/m_{\omega}^2$. This yields the relation between the real and the effective chemical potentials
\begin{equation} \label{eq:effchempot}
	\mu = \nu + G_\omega \tilde{n}(T,\nu).
\end{equation}
Moreover, the corresponding thermodynamic potential for nonzero vector coupling is given by
\begin{equation} \label{eq:OmegaVector}
	\Omega(T,\mu) = -P(T,\mu) = \tilde{\Omega}(T,\nu) - \frac{(\mu-\nu)^2}{2 G_{\omega}}
\end{equation}

We note,  that for a first order phase transition,  this procedure cannot be used to identify the  critical value of thermal parameters, owing to a flattening of the potential, as a function of the order parameter.

\subsection{Net-baryon-number cumulants}
Fluctuations of conserved charges, in particular of the  net baryon number, are valuable probes of critical phenomena and can be used to identify the location of the  phase boundary of chiral symmetry restoration. The fluctuations are characterized by cumulants, that can be obtained  directly from the partition function. The  $n^{\mathrm{th}}$-order baryon $\chi_B^n$,  and quark  $\chi_q^n$ number cumulants   are
\begin{equation}\label{eq:chidef}
	3^n \chi_B^n = \chi_q^n = \frac{1}{T^{4-n}}\frac{\partial^n \Omega(T,\mu_q)}{\partial \mu_q^n}.
\end{equation}
At vanishing vector coupling, the first order cumulant  $\chi_B^1$, the net baryon density, can be obtained directly from  the flow of $\partial \Omega_k / \partial\mu$, while  the second order cumulant  $\chi_B^2$ is given by
\begin{align}\label{2nd}
	\chi_B^2 = \frac{1}{9T^2} \bigg( &\frac{\partial^2 \Omega_{k=0}}{\partial\mu_q^2 } \nonumber \\
	&- \sum_{i,j} \frac{\partial^2 \Omega_{k=0}}{\partial\mu_q \partial \varphi_i}
	\left(\frac{\partial^2 \Omega_{k=0}}{\partial\varphi \partial \varphi}\right)^{-1}_{ij}
	\frac{\partial^2 \Omega_{k=0}}{\partial\mu_q \partial \varphi_j} \bigg),
\end{align}
where $\varphi=(\sigma,\Phi,\bar{\Phi})$. The derivatives on the right hand side of Eq. (\ref{2nd}) are determined by  the solution of flow equations introduced in Appendix \ref{sec:flows}.
Higher order cumulants are then computed  using  numerical differentiation of  $\chi_B^2$.

For  nonvanishing vector coupling, the baryon number cumulants  can be expressed in terms of the corresponding  cumulants  $\tilde{\chi}_B^n$ for vanishing  vector coupling.  This is done using Eqs.~\eqref{eq:effchempot} and \eqref{eq:OmegaVector}.
Based on Eq.~\eqref{eq:effchempot} one can write
\begin{equation}\label{eq:Dchempot}
	 \frac{d\nu}{d\mu} = \frac{1}{1+9T^2 G_{\omega}\tilde{\chi}_B^2(T,\nu)},
\end{equation}
which using Eq. (\ref{eq:OmegaVector}), leads to the following relations
\begin{align} \label{eq:transrules}
\begin{split}
	\chi_B^1 &= \tilde{\chi}_B^1, \\
	\chi_B^2 &=  \frac{\tilde{\chi}_B^2}{1+9T^2 G_{\omega}\tilde{\chi}_B^2(T,\nu)}, \\
	\chi_B^3 &=  \frac{\tilde{\chi}_B^3}{\left(1+9T^2 G_{\omega}\tilde{\chi}_B^2\right)^3}, \\
	\chi_B^4 &=  \frac{\tilde{\chi}_B^4}{\left(1+9T^2 G_{\omega}\tilde{\chi}_B^2\right)^4}
	-\frac{27T^2 G_{\omega} \left(\tilde{\chi}_B^3\right)^2}{\left(1+9T^2 G_{\omega}\tilde{\chi}_B^2\right)^5}.
\end{split}
\end{align}
Here the cumulants $\tilde{\chi}_B^k$ are  functions of  temperature and the reduced chemical potential $\nu$.
Using these relations, we can compute the influence of the vector interaction on the baryon number cumulants, using the RG flow obtained for vanishing vector coupling, $G_\omega=0$.

\section{Net-baryon-number cumulants and  the phase boundary \label{sec:phaseboundary}}

In the previous  section we have introduced the  method employed to compute  cumulants of   net baryon number in an effective chiral model in the presence of vector interactions.
In the following, we discuss the properties of these cumulants near,  and at the chiral phase boundary,  at finite temperature and chemical potential.

The chiral Lagrangian introduced above, shares the chiral critical properties with QCD. In particular,  at moderate values of the chemical potential, the PQM model exhibits a chiral transition belonging to the $O(4)$ universality class \cite{Pisarski:1983ms,Schaefer:2006ds}. For larger values of $\mu$, it reveals a $Z(2)$ critical endpoint, followed  by a first order  phase transition \cite{Schaefer:2006ds}. Consequently, the PQM model embodies the generic  phase structure expected for QCD,  with the universal $O(4)$ and $Z(2)$ criticality  encoded in the scaling functions. Furthermore, due to the coupling of the quarks to the background gluon fields, the PQM model incorporates  "statistical confinement", i.e., the suppression  of quark and diquark degrees of freedom in the low temperature,  chirally broken phase~\cite{Fukushima:2003fw,Skokov:2010sf}. Consequently,  by studying fluctuations of conserved charges in the PQM model, one  can explore the influence of chiral symmetry restoration and of "statistical confinement"  on the cumulants  in  different sections   of  the chiral phase boundary. This study is  of particular interest in the context of heavy-ion collisions, where  cumulants of conserved charges are expected to provide a characteristic signature for the QCD phase boundary and for the conjectured critical endpoint~\cite{Stephanov:1998dy,Stephanov:1999zu,Asakawa:2000wh,Jeon:2000wg,Friman:2011pf,Ejiri:2005wq,Karsch:2010ck}.

\newpage
\subsection{Criticality of net-baryon-number cummulants in the \texorpdfstring{$(T,\mu)$}{(T,mu)} plane}

Generalized susceptibilities of conserved charges, which are given by higher-order derivatives of the thermodynamic pressure with respect to the corresponding chemical potentials, may exhibit a nonmonotonic dependence on the thermodynamic parameters. This is particularly the case  in the vicinity of phase boundary and the CEP. In the critical region of the chiral transition, the strength of the  fluctuations and the sign of the susceptibilities are by and large determined by the singular part of the free energy, which is encoded in the universal scaling functions, common  to QCD and the PQM model. Thus, generic structures of the susceptibilities and  relations  between  them  near the phase boundary, can also be studied in the PQM model. Of particular interest is  the behavior of susceptibilities along the $O(4)$ crossover line, and their modification as the CEP is approached. In the model calculations, the  position  of the CEP depends on the strength of the vector interaction. Thus, by changing the vector coupling $g_\omega$, one can assess the dependence of critical fluctuations in different sectors of the phase boundary on the location of the CEP.

\begin{figure}[tb]
   \includegraphics[width=0.99\linewidth]{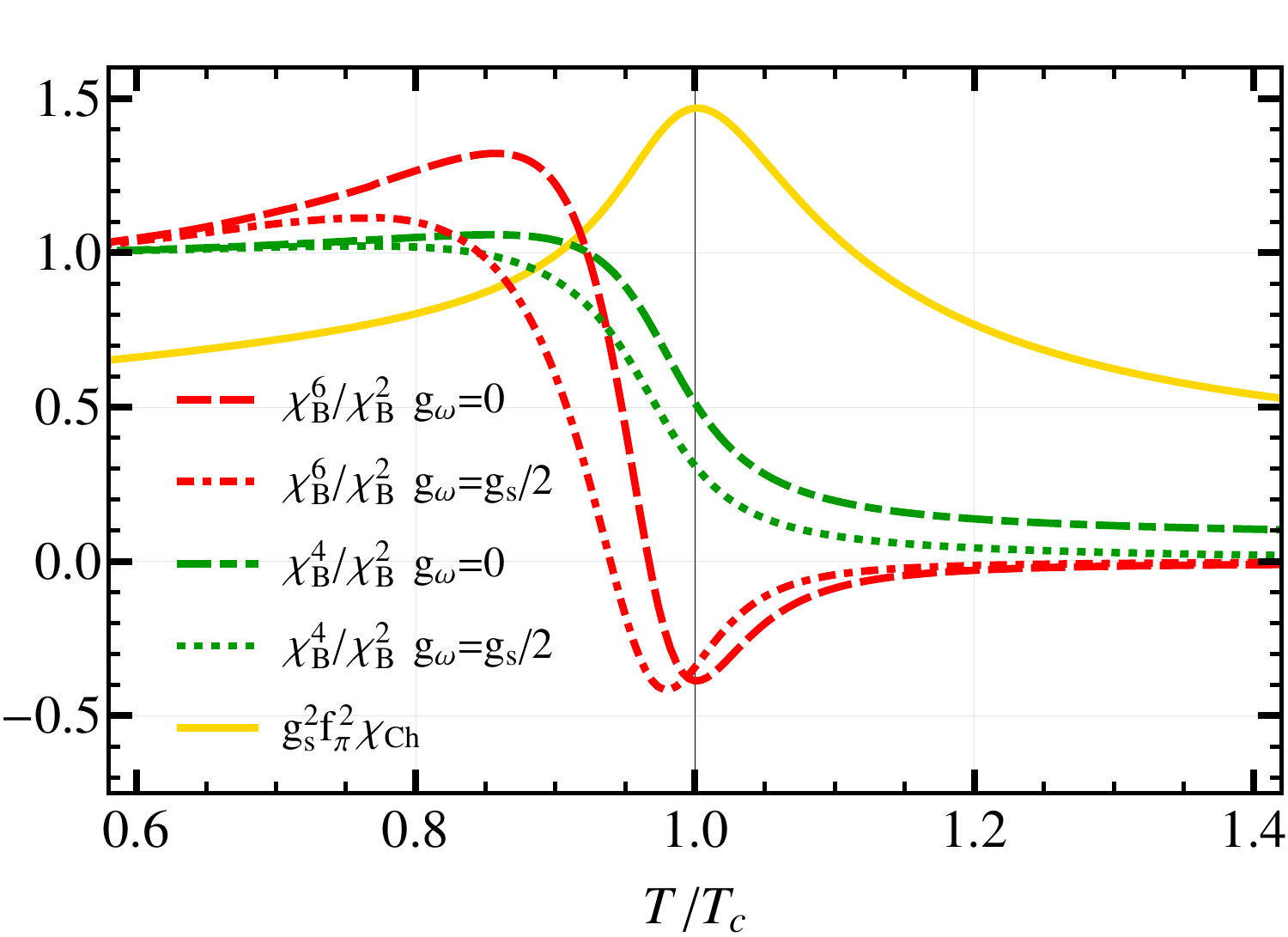}
   \caption{The temperature dependence of ratios of net-baryon-number   cumulants, $\chi_B^4/\chi_B^2$ and  $\chi_B^6/\chi_B^2$ in the PQM model for vanishing and nonzero vector coupling $g_\omega$. Also shown is the chiral susceptibility, $\chi_{ch}(T)$.
 The vertical line indicates the location of the chiral crossover temperature. \label{fig:one}}
\end{figure}

For nonvanishing light-quark masses, the chiral symmetry is explicitly broken, which implies that at finite temperatures and moderate values of the  baryon chemical potential, the system exhibits  a chiral  crossover transition. Thus,  at small $\mu_q$,  the fluctuations of conserved charges remain finite. However, because QCD matter at the physical values of the $u$ and $d$ quark masses is within the critical region of the second order phase transition, the fluctuations are still influenced by $O(4)$ criticality~\cite{Ejiri:2009ac,Kaczmarek:2011zz,Ding:2013lfa,Ding:2015pmg,Hatta:2002sj}. At physical quark masses, a genuine phase transition in QCD and the associated  singular behavior of fluctuations are only expected at the  conjectured  CEP and along the line of the first-order phase transition.

\begin{figure*}[tb]
   \includegraphics[width=0.49\linewidth]{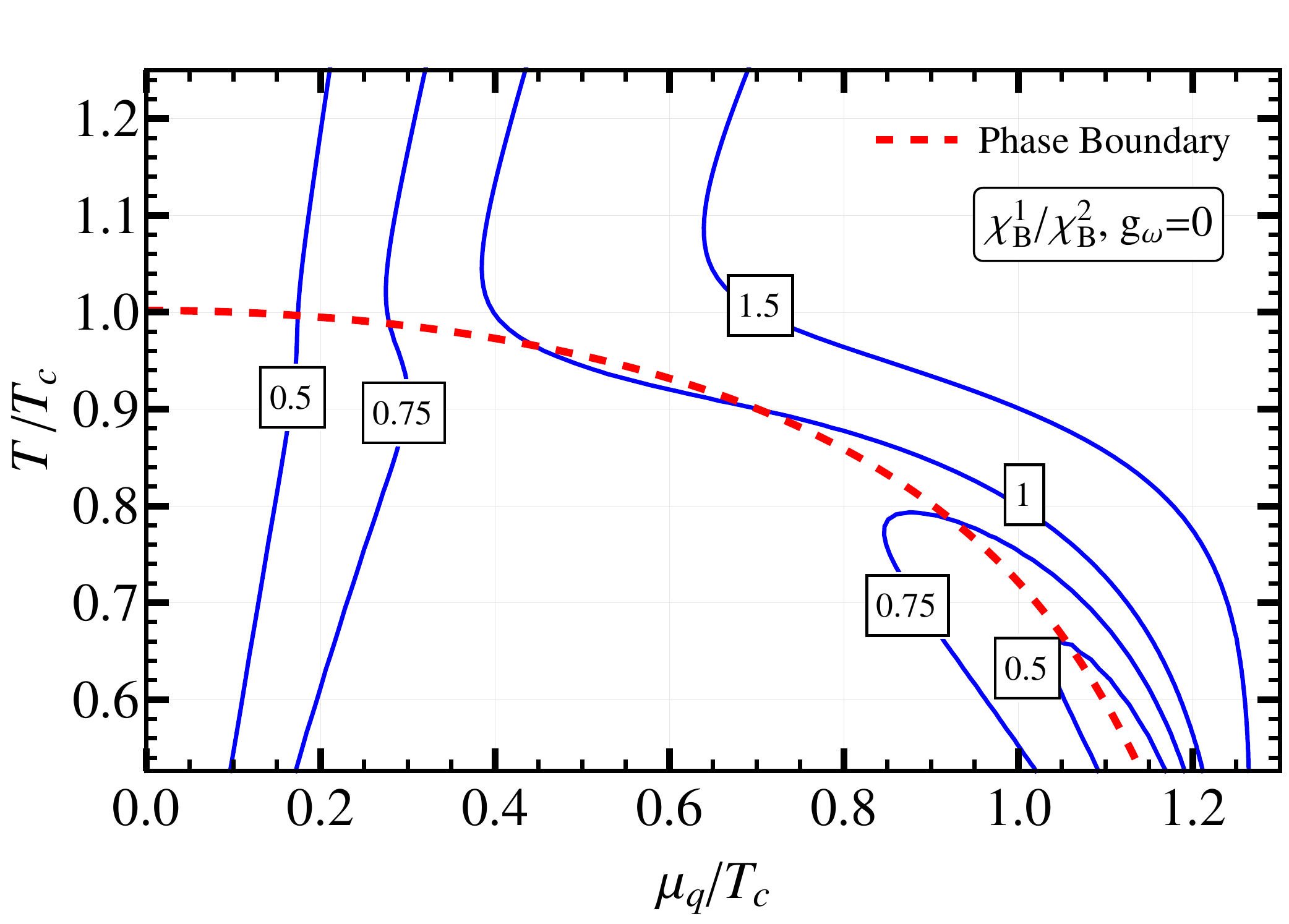}
   \includegraphics[width=0.49\linewidth]{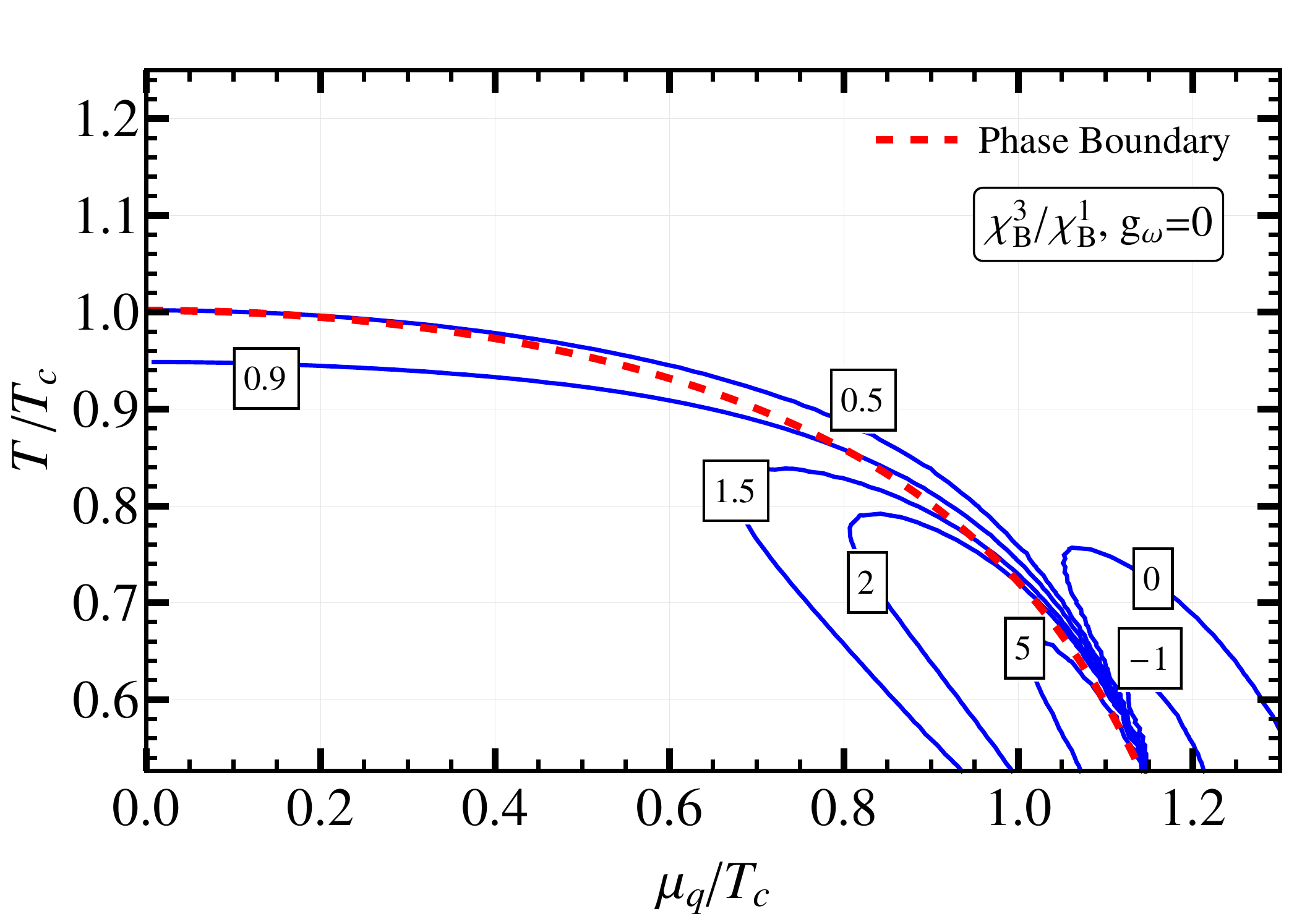}
   \caption{Contour plots of the ratios  $\chi_B^1/\chi_B^2$  and $\chi_B^3/\chi_B^1$   in the $(T,\mu)$-plane, computed in the PQM model. The broken lines indicate the location of the chiral crossover phase boundary. \label{fig:two}}
\end{figure*}

\begin{figure*}[tb]
   \includegraphics[width=0.49\linewidth]{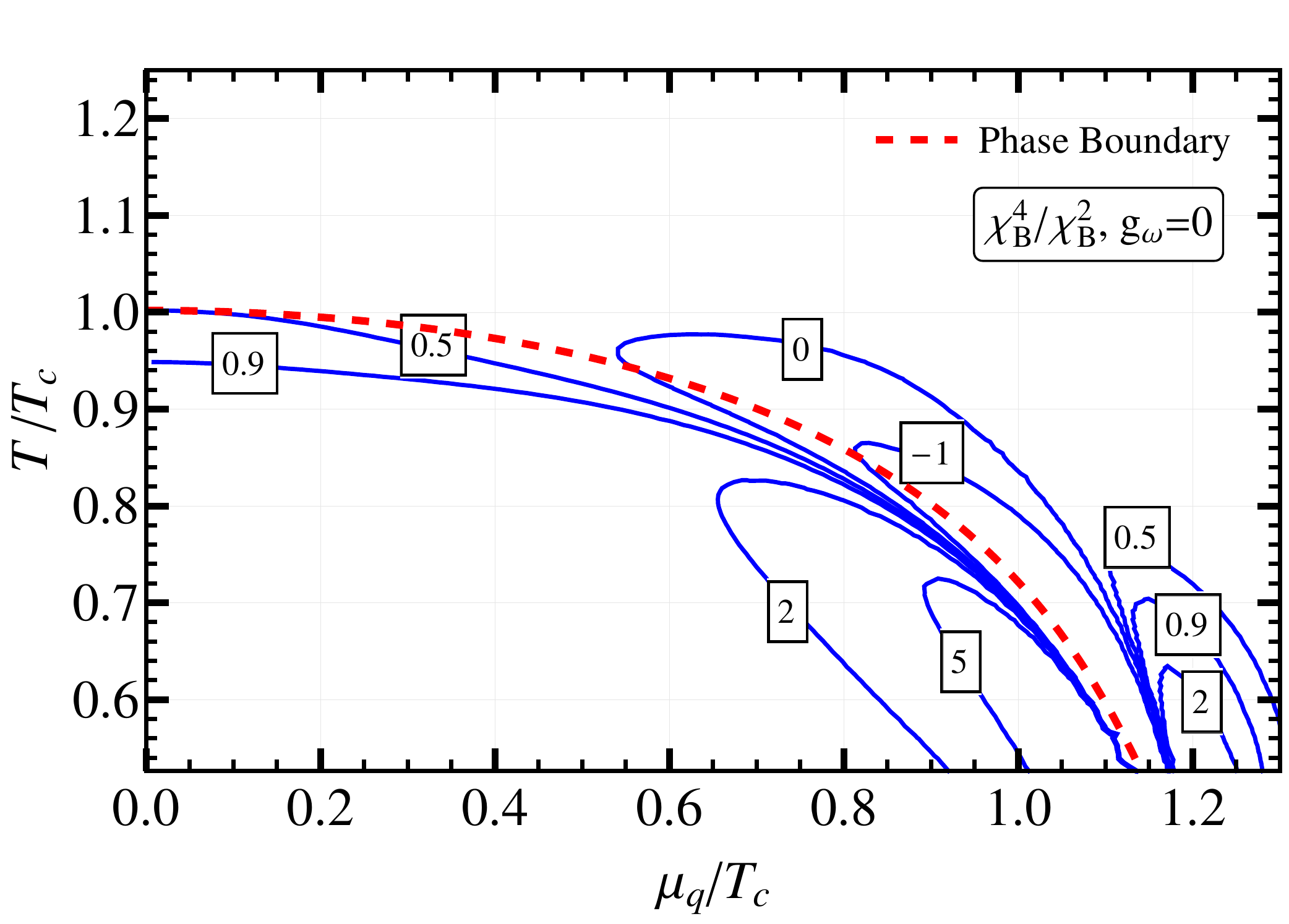}
	\includegraphics[width=0.49\linewidth]{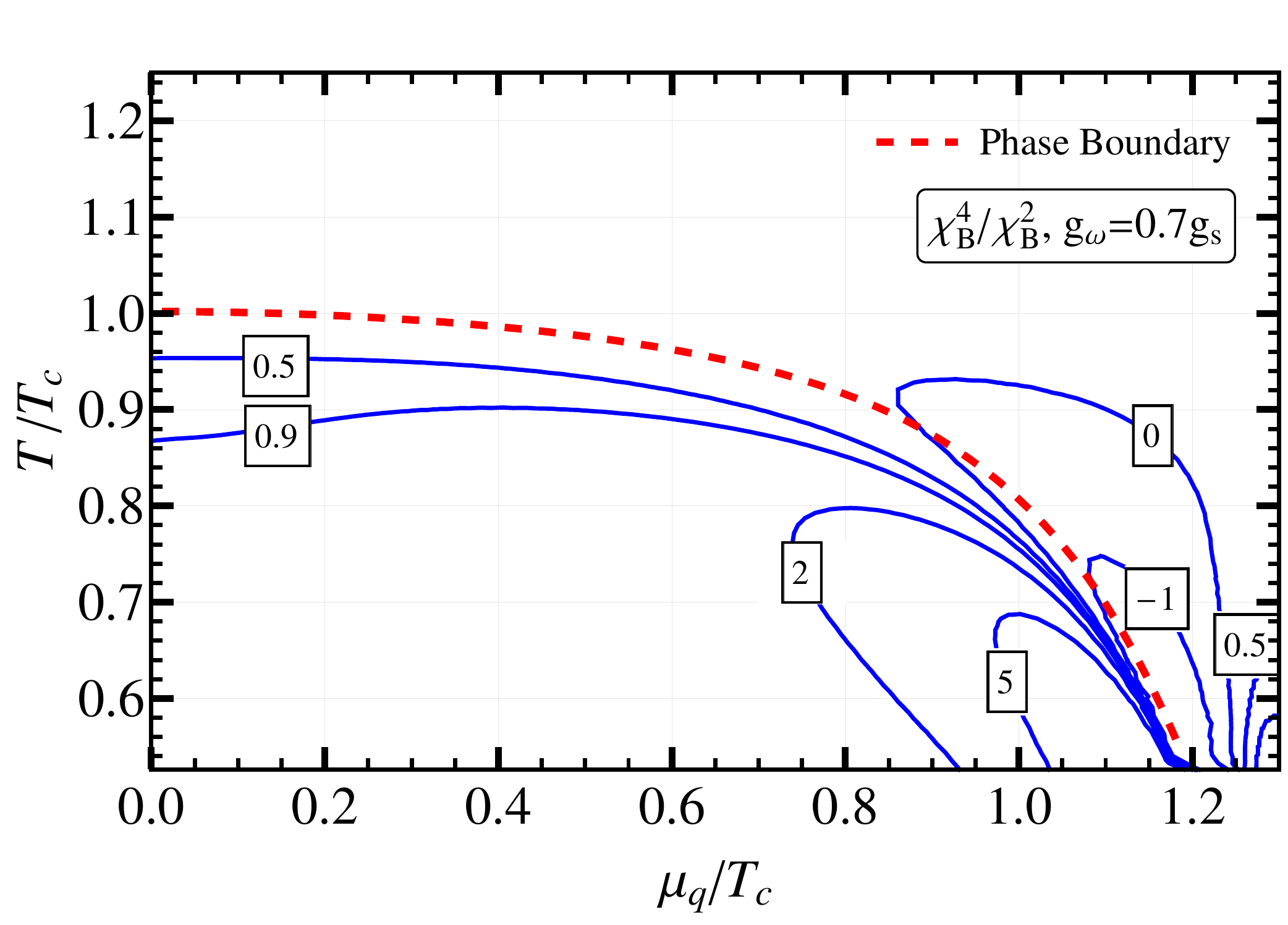}
   \caption{ Contour plots  of the kurtosis ratio,  $\chi_B^4/\chi_B^2$,
   in the $(T,\mu_q)$-plane. The left-hand figure corresponds to PQM model results  obtained with  vanishing vector coupling    $g_{\omega}=0$, while the right-hand one shows  results obtained with  $g_\omega=0.7 g_s$.
\label{fig:three}}
\end{figure*}

To explore the phase diagram with net-baryon-number fluctuations, we need to know the qualitative dependence of net-baryon-number cumulants on the temperature and baryon chemical potential. In Fig.~\ref{fig:one} we show the  $T$-dependence at $\mu_q=0$,  and in Figs.~\ref{fig:two} and ~\ref{fig:three},  the  contour plots in the $(T,\mu_q)$-plane of ratios of net-baryon-number susceptibilities. We  focus on the following ratios of net-baryon-number susceptibilities:

\begin{align}
	 \chi_B^{n,m}&=\frac{\chi_B^{n}(T,\mu_B)}{\chi_B^{m}(T,\mu_B)}, \\
	 \kappa\sigma^2&=\frac{\chi_B^{4}(T,\mu_B)}{\chi_B^{2}(T,\mu_B)}, ~~\frac{S_B\sigma^3}{M}=\frac{\chi_B^{3}(T,\mu_B)}{\chi_B^{1}(T,\mu_B)},
\label{def}
\end{align}
where $M$ is the mean, $\sigma$ the variance, $S_B$ the skewness and $\kappa$ the kurtosis of the net-baryon-number distribution. The statistical descriptors are related to the susceptibilities through $M=V T^3 \chi_B^1$, $\sigma^2=V T^3 \chi_B^2$ etc..

At vanishing chemical potential, all odd susceptibilities of net baryon number vanish, owing to the baryon-antibaryon symmetry. In addition, in the $O(4)$ universality class, the second and fourth order cumulants remain finite at the phase transition temperature  at $\mu_q=0$ even in the chiral limit, implying that only sixth and higher order susceptibilities  diverge. Thus, for physical quark masses, only higher order cumulants,   $\chi_B^n$ with $n>4$, can exhibit $O(4)$ criticality at $\mu_q=0$~\cite{Friman:2011pf}. A further consequence of the baryon-antibaryon symmetry is the equality of the ratios
\begin{equation}
\chi_B^{2m-1,2n-1}=\chi_B^{2m,2n}
\label{eq:odd-even}
\end{equation}
for any integer $m$ and $n\,\geq 1$ at $\mu_q=0$. For $\chi_B^{3,1}$ $\chi_B^{4,2}$, the equality at small $\mu_q$ can also be confirmed by a direct comparison of the right panel of Fig.~\ref{fig:two} with the left panel of Fig.~\ref{fig:three}.

At finite net baryon density the singularity at the $O(4)$ critical line is stronger than at $\mu_q=0$. Thus, in this case the third-order cumulant and all higher-order ones, diverge at the $O(4)$ line. The second order cumulant $\chi_B^2$  remains finite, and only diverges at the tricritical point for vanishing quark masses, and at the CEP for nonzero quark masses.


In Fig.~\ref{fig:one} we show  results on the temperature dependence of several ratios $\chi_B^{n,m}$ of net-baryon-number susceptibilities,    together with the variance  of the chiral condensate,  $\chi_{ch}$.
The location of the maximum of the  chiral susceptibility, $\chi_{ch}$, defines  the pseudo-critical temperature,  $T_c$.

As shown in Fig.~\ref{fig:one}, the  kurtosis $\kappa\sigma^2$  exhibits a rapid drop by approximately an order of magnitude at the phase boundary. This strong reduction of the kurtosis is attributed to the deconfinement transition, where the degrees of freedom carrying baryon number change from baryons to quarks \cite{Karsch:2005ps,Ejiri:2005wq}. A qualitative understandning of this issue is obtained in the Boltzmann approximation, where the baryon contribution to the thermodynamic pressure is of the form $P_B(T,\mu_B)\simeq F(T)\cosh(B\mu_B/T)$. Here $B$ is the  baryon number of the relevant degrees of freedom and $F(T)$ a function of the corresponding excitation spectrum. At low temperatures, the degrees of freedom are baryons, with $B=\pm 1$, while at high temperatures they are quarks with $B=\pm 1/3$. Consequently, the kurtosis ratio, $\kappa\sigma^2=\chi_B^4/\chi_B^2$, is approximately proportional to the square of the baryon number of the relevant degrees of freedom \cite{Karsch:2005ps,Ejiri:2005wq}. This  clarifies the cause for the rapid  change of  $\kappa\sigma^2$ at the crossover transition from $\simeq 1$ to $\simeq 1/9$, seen  in Fig.~\ref{fig:one}. In the limit where the characteristic  mass is small compared to the temperature, the kurtosis ratio is modified by quantum statistics,  to {$\kappa\sigma^2=(1/9)(6/\pi^2)$}.
We note,  that the kurtosis ratio is independent of the  mass spectrum as well as of any kinematic cuts, as long as the Boltzmann limit remains valid.

The ratio $\chi_B^{6,2}$, also shown in Fig.~\ref{fig:one}, exhibits more structure at the chiral transition. The  observed  characteristic temperature dependence of this ratio, in particular  with a region of negative values at $T\simeq T_c$, is a consequence of the residual chiral $O(4)$ criticality~\cite{Friman:2011pf}. In the absence of chiral critical fluctuations, $\chi_B^{6,2}$  would show  a similar behavior as $\kappa\sigma^2$, with a smooth reduction from unity at low temperatures towards zero above the deconfinemnt transition. The distinctive behavior of the $\chi_B^{6,2}$ and $\chi_B^{4,2}$ ratios  was already obtained in the PQM model within the FRG approach~\cite{Friman:2011pf}, and agrees qualitatively with   LQCD  results~\cite{Bazavov:2017dus}.  Thus, the PQM model correctly captures the physics of QCD related to deconfinement and to the critical dynamics at the chiral transition. The negative values of $\chi_B^{6,2}$ at the chiral crossover  were proposed as a signature for partial restoration of chiral symmetry in heavy ion collisions \cite{Friman:2011pf}.
\newpage
As shown in Fig.~\ref{fig:one}, in the presence of repulsive interactions,  $\chi_B^{6,2}$  and the kurtosis  ratios  are modified.  The vector interaction leads to  a downward shift in temperature of the ratios relative to the phase boundary, as well as, a suppression of the sixth order susceptibility in the temperature range below $T_c$. Nevertheless, the qualitative form of the ratios is preserved. In particular, the characteristic structure, where the sixth order cumulant is negative in a range of temperatures near $T_c$ owing to $O(4)$ criticality, is not eliminated by the repulsive interaction.

In  Fig.~\ref{fig:two} we show contour plots of the ratios $\chi_B^{1,2}$ and $\chi_B^{3,1}$  in the $(T,\mu_q)$-plane. As noted above, all odd cumulants vanish at $\mu_q=0$, owing to baryon-antibaryon symmetry. Consequently, $\chi_B^{1,2} |_{\mu_q=0}=0$ for any $T$, while the ratio
$\chi_B^{3,1} |_{\mu_q=0}$ is nonvanishing. At  low  and high $T$, relative to  $T_c$, this ratio is consistent with  unity and {$2/(3\pi^2)$}, respectively,  as expected since $\chi_B^{3,1} |_{\mu_q=0}=\chi_B^{4,2} |_{\mu_q=0}$. As  indicated in Figs.~\ref{fig:one} and \ref{fig:two}, the ratio  $\chi_B^{3,1} $ decreases with temperature, and depends weakly on the chemical potential. Moreover, Fig. \ref{fig:two}, shows that for $\mu_q<T$,  $\chi_B^{1,2}$ increases with $\mu_q$, and is only  weakly dependent on the temperature. Thus, the ratio $\chi_B^{3,1}$ can be used as a measure of the temperature, while $\chi_B^{1,2}$ provides a gauge  of the chemical potential.


At small $\mu_q/T$, the properties of the first four susceptibilities,  $\chi_B^n$ with $n=1,..,4$,  and consequently their ratios near the chiral crossover are dominantly affected by the coupling of the quarks to the Polyakov loop, and the resulting statistical confinement. The critical chiral dynamics, i.e.  the  $O(4)$ and $Z(2)$ criticality at the chiral crossover transition and at the CEP, respectively, unfolds at larger $\mu_q/T$.  Near the CEP, there is a strong variation of the cumulants  with $T$ and $\mu_q$,  which   increases  with the order of cumulants. As shown in  Figs.~\ref{fig:two} and  \ref{fig:three}, the qualitative behavior of the cumulant ratios on lines going to the CEP depends strongly on the direction from which  the CEP is approached. This behavior   is governed by the critical properties encoded  in the  $O(4)$ and $Z(2)$ universal scaling functions.

The influence of the CEP on the characteristics of the various cumulant ratios can be studied by varying the value of the vector coupling, thus changing the position of the CEP in the $(T,\mu_q)$-plane. This is illustrated  in Fig. 3 for the kurtosis ratio. A comparison of the results shown in  the left and right contour plots shows that with increasing $g_\omega$, the curvature of the phase boundary is reduced and the position of the CEP is shifted to  lower $T$ and larger $\mu_q$ \cite{Kitazawa:2002bc}. As is clearly seen, when comparing the left and right plots of Fig.~\ref{fig:three}, there is  a corresponding shift of the contours of   $\chi_B^{4,2}$. These results clearly illustrate the influence of the critical endpoint  on the fluctuation observables. A shift of the CEP to larger net baryon density, suppresses the magnitude of the net-baryon-number susceptibilities at a given  $T$ and $\mu_q$.

\subsection{Net-baryon-cumulant ratios and freeze-out in heavy ion collisions\label{sec:FO}}

\begin{figure*}[tb]
   \includegraphics[width=0.49\linewidth]{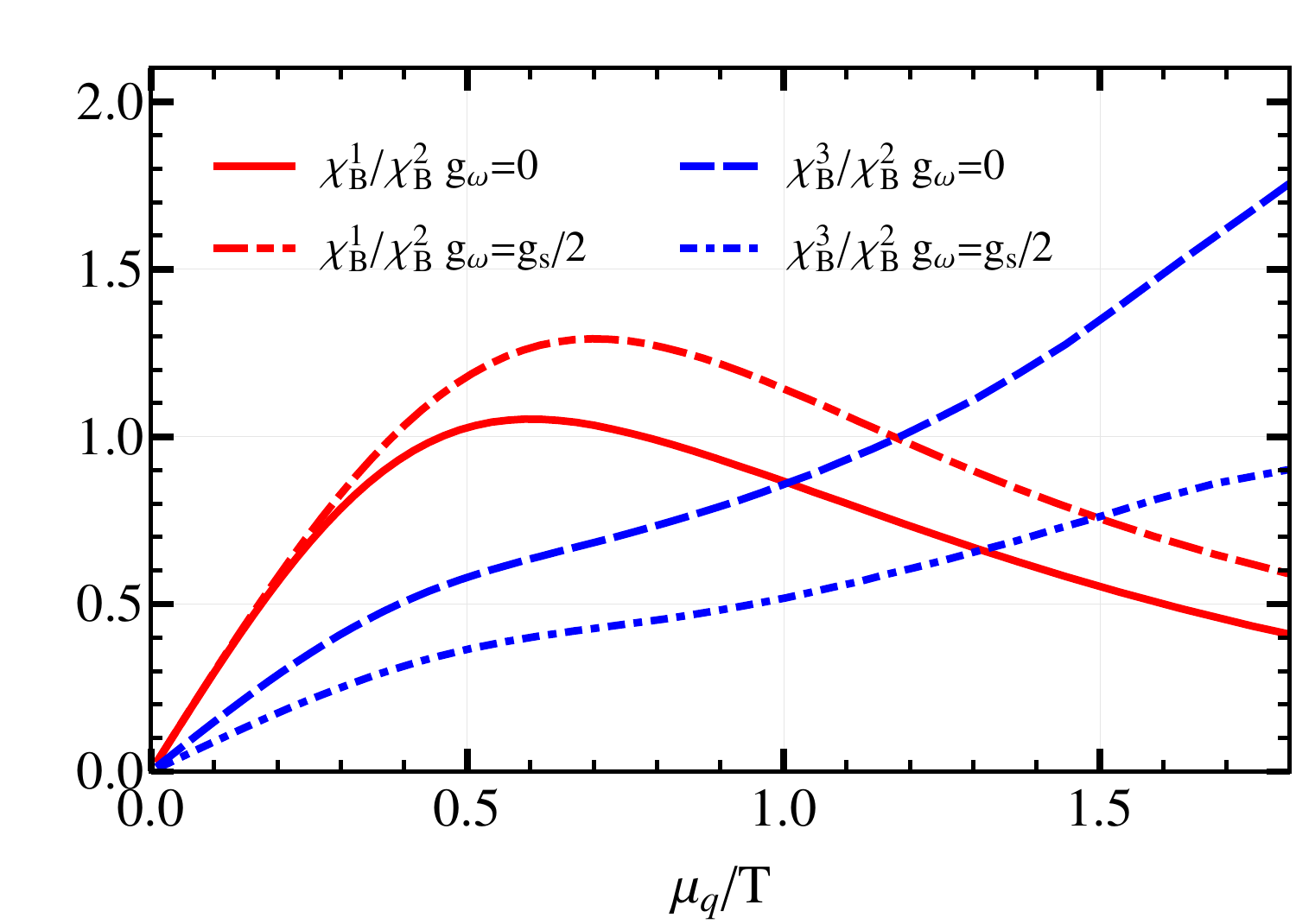}
	\includegraphics[width=0.49\linewidth]{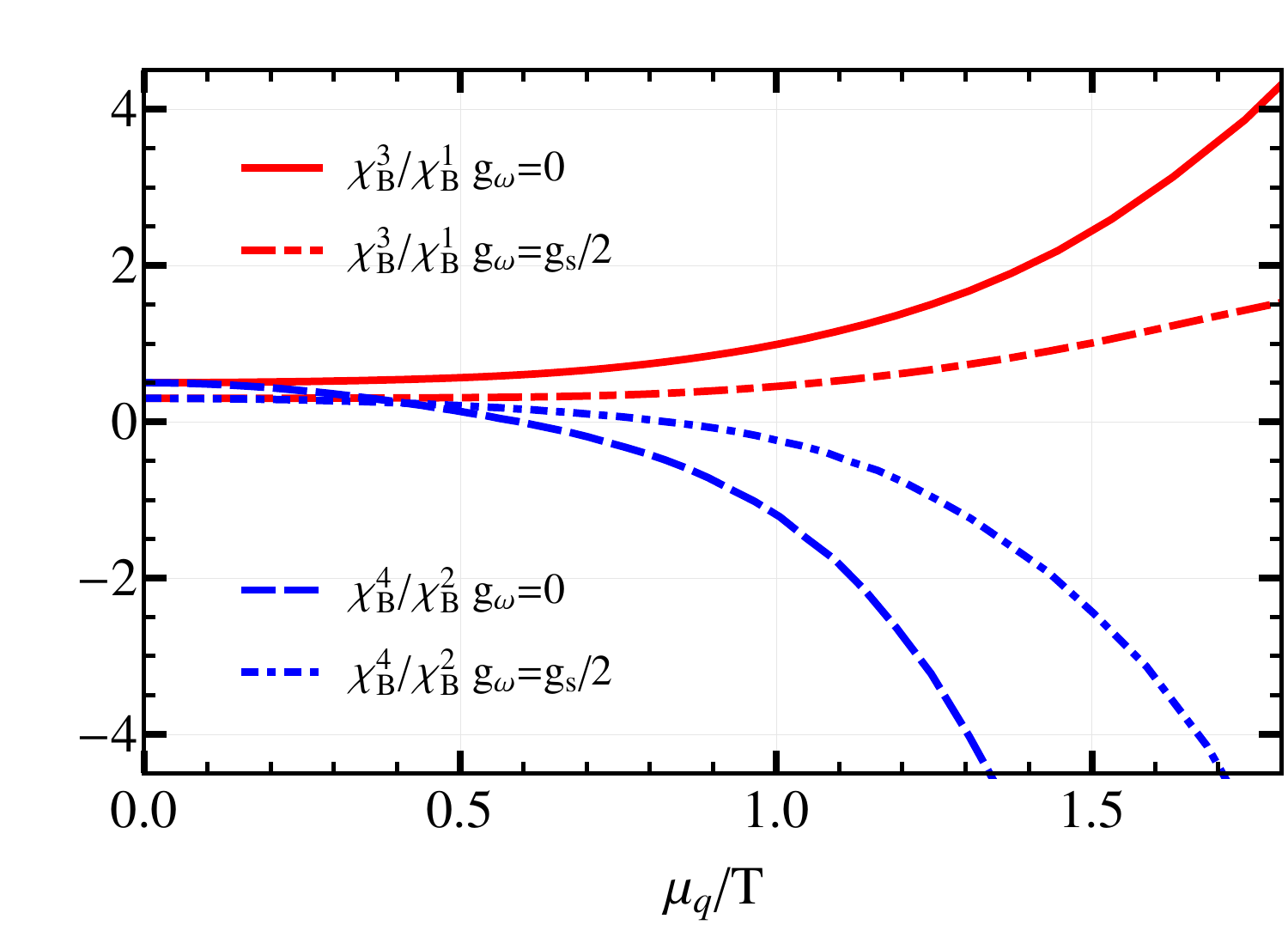}
   \caption{Ratios of cumulants of net-baryon-number fluctuations  in the PQM model computed on  the chiral phase boundary with and without  vector repuslion. The solid and long-dashed lines correspond to $g_\omega=0$, and the dash-dotted lines  to $g_\omega=g_s/2$.
  \label{fig:four}}
\end{figure*}

In heavy-ion collisions, the thermal fireball  formed  in the   quark-gluon plasma phase
undergoes expansion and passes through the QCD phase boundary  at some point   $(T_c,\mu_c)$ point, which depends on  the collision energy,  $\sqrt s$. Analysis of ratios of particle multiplicities indicate that at high beam energies (small values of $\mu_q/T$), the freeze-out occurs at, or just below  the phase boundary.  Thus, the beam energy dependence of net-baryon-number  susceptibilities  can provide insight into the structure of the QCD phase diagram and information on the existence and location  of the  CEP. Consequently, it is of phenomenological interest to compute the    properties of  fluctuations of conserved  charges along the chiral  phase boundary. Since there,  the  critical structure and the relations between different susceptibilities are governed by the universal scaling functions, the   generic behavior of  ratios of net-baryon-number susceptibilities  can be explored  also in  model calculations.

In Fig.~\ref{fig:four} we show  $\chi_B^{1,2}$, as well as skewness  and  kurtosis ratios   of  net baryon number, obtained in the PQM model  along the chiral phase boundary,  with and without vector repulsion. In the former case, the position of the CEP  is  shifted to smaller $T$ and larger $\mu_q$.

The ratio $\chi_B^{1,2}$ exhibits  a  maximum along the phase boundary. For small $(\mu_q/T)_c$ this ratio is well approximated by $\tanh(3\mu_q/T)$, while after reaching a maximum,  it decreases as the CEP is approached. The cumulants $\chi_B^1$ and  $\chi_B^2$ remain finite  along the $O(4)$ line. However, at the CEP the variance of the net-baryon-number fluctuations diverges. This is the cause for the observed decrease of  $\chi_B^{1,2}$  at larger $(\mu_q/T)_c$. For nonzero vector repulsion, this reduction is weakened, owing to the shift of the CEP to lower temperature. We note, that for  $(\mu_q/T)_c<0.5$, the $\chi_B^{1,2}$ ratio is hardly  modified by the repulsive interactions. This can be traced back to a cancellation between the suppression of $\chi_B^2$  at nonzero  $g_\omega$ (\ref{eq:transrules}) and the shift of the  chemical potential (\ref{eq:effchempot}); both axes are scaled by the factor $(1+9T^2 G_{\omega}\tilde{\chi}_B^2(T,\nu))$,  for small $\mu_q$. On the other hand, the ratio $\chi_B^3/\chi_B^2$ is  \textit{reduced} by the factor $(1+9T^2 G_{\omega}\tilde{\chi}_B^2(T,\nu))$, for nonvanishing vector interaction. Hence, this ratio is reduced at all values of $\mu_q/T$.

As seen in Fig.~\ref{fig:four}, the  skewness ratio, $\chi_B^{3,1}$,  is an increasing function  of $(\mu_q/T)_c$ along the phase boundary and  at the CEP it  diverges~\footnote{Since the CEP is a singular point, the corresponding value of $\chi_B^3$ depends on how the limit is taken. Owing to the curvature of the phase boundary, the CEP is approached slightly from below, which implies that $\chi_B^3$ diverges to plus infinity.}. There is also a  clear suppression of  this ratio along the phase boundary due to repulsive interactions, as seen also for $\chi_B^{3,2}$.  At small  $(\mu_q/T)_c=0$, the skewness and kurtosis ratios are  equal to each other,  and differ very little up to  $(\mu_q/T)_c\simeq 0.5$. For larger $(\mu_q/T)_c$,  the kurtosis is decreasing and skewness increasing along the phase boundary, in agreement  with recent LGT results~\cite{Karsch:2016yzt}. This behavior also  reflects their properties near the critical point.  As the CEP is approached along  the phase boundary, the kurtosis diverges to minus infinity, while the skewness diverges to plus infinity, as noted above.
Thus, both ratios becomes less singular as $g_\omega$ is increased, i.e., as the CEP is shifted to larger $\mu_q$.

The characteristics of the various net-baryon-number susceptibilities  on the phase boundary,  shown in Fig.~\ref{fig:four},  are expected to be similar in QCD.  This is because, they are, by and large, determined by $O(4)$ critical fluctuations and by "confinement", which are both common to QCD and the PQM model. This opens the possibility to verify directly,  if these features  of criticality  are also reflected in the data on  net-proton-number fluctuations obtained in heavy-ion collisions by the STAR Collaboration at several RHIC energies~\cite{Adamczyk:2013dal,Luo:2015ewa,Luo:2015doi}.

\begin{figure*}[tb]
   \includegraphics[width=0.49\linewidth]{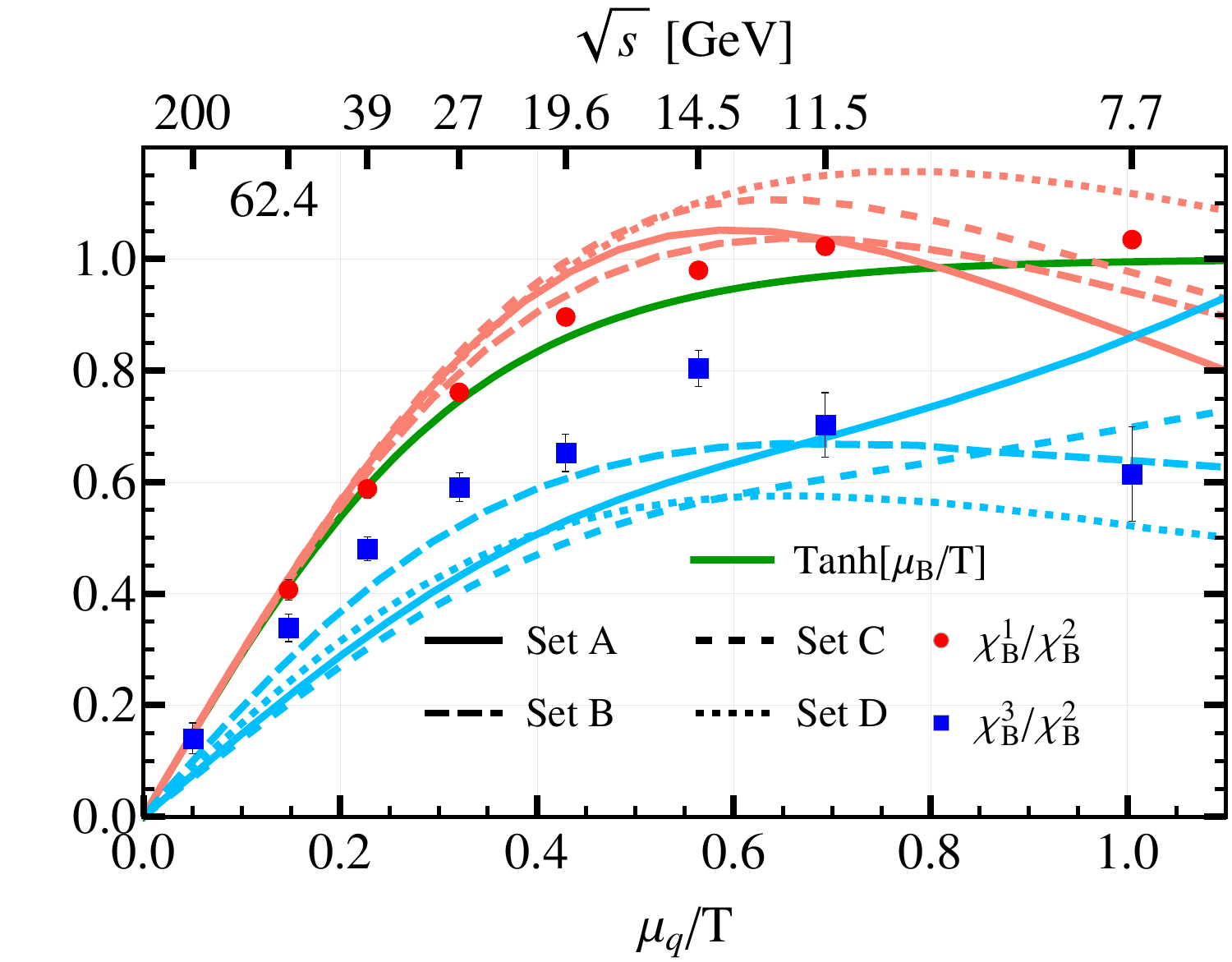}
	\includegraphics[width=0.49\linewidth]{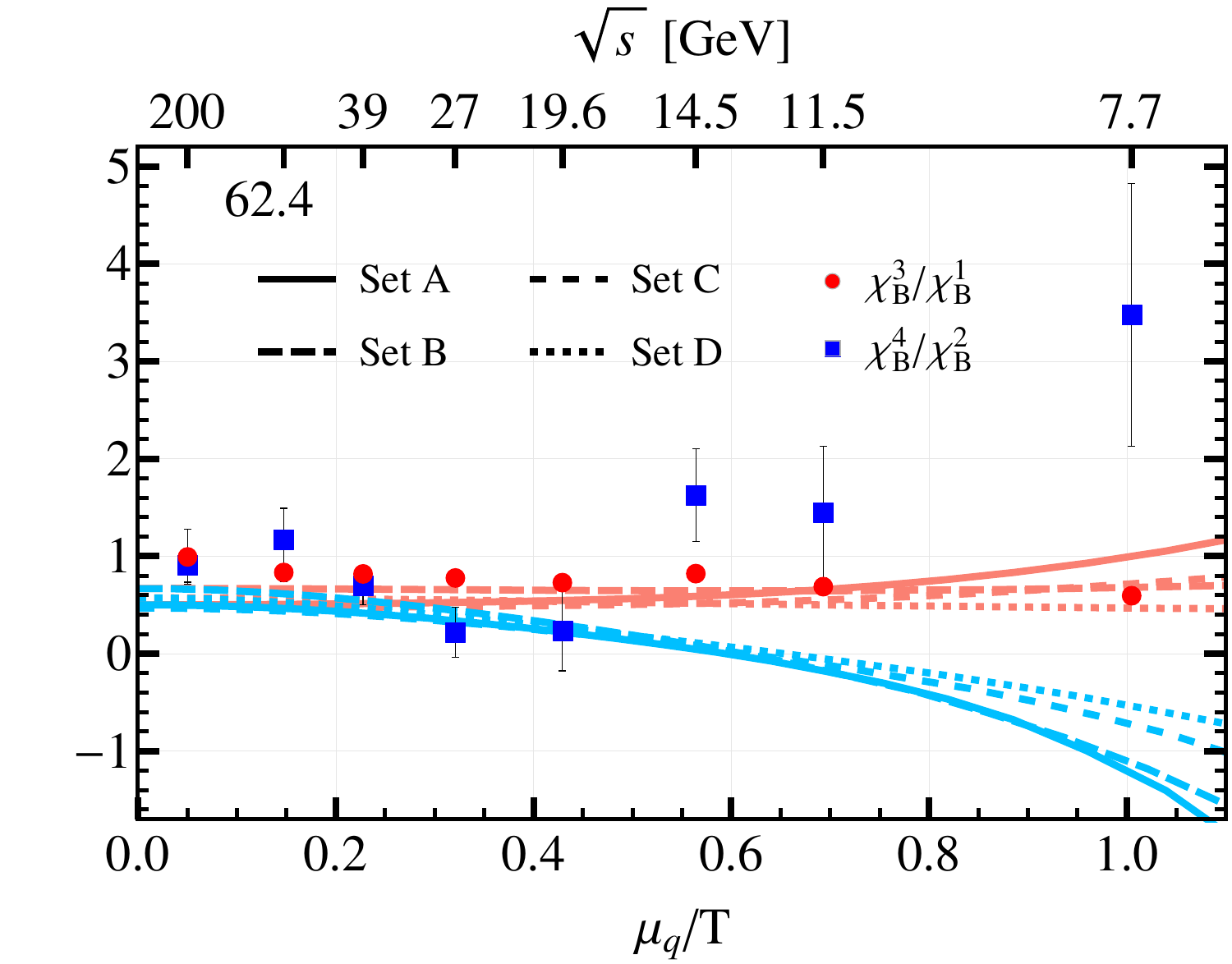}
   \caption{
   Ratios of cumulants of net-baryon-number fluctuations  in the PQM model, computed along the chiral phase boundary,  for four sets of model parameters (for details see Appendix \ref{sec:initc}). Also shown are the preliminary STAR data~\cite{Luo:2015ewa,Luo:2015doi}, assuming the relation between the ratio $(\mu_q/T)$  and the collision energy obtained by analysing the chemical freeze-out conditions \cite{Andronic:2005yp,Andronic:2008gu,Cleymans:2005xv}. The green full line in the left-hand  figure shows the baseline result,  $\chi_B^{1,2}= \tanh(3\mu_q/T)$.
 \label{fig:five}}
\end{figure*}

Clearly,  a direct comparison of model results with data has to be taken with  caution.  Although, the model provides a viable description of the dynamics that  drives the system towards chiral symmetry restoration,  the spectrum of hadronic degrees of freedom in the low temperature phase is incomplete.  Moreover, net-baryon-number fluctuations are  in nucleus-nucleus collision experiments quantified by the net proton number. It has been extensively discussed,  to what extent are net-proton-number fluctuations accurate proxyies for those of the net baryon number~\cite{Bzdak:2012ab,Bzdak:2012an,Kitazawa:2011wh,Kitazawa:2012at}.         Furthermore, there are   kinematical  cuts,  imposed  on the STAR data on the cumulants of net proton number, which are not accounted for in the model results.  However, these differences are, to a large extent, eliminated  by considering ratios of susceptibilities.
This assumption is supported by the behavior of the ratio $\chi_B^{1,2}$,   which, in spite of the differences discussed above, is well approximated by $\tanh(3\mu_q/T)$ (see Fig. \ref{fig:five}) in LQCD, in the PQM model as well as in the STAR data. The fact that $\chi_B^{1,2}$ is well approximated by this functional form indicates that in the transition region, the effective degrees of freedom with nonvanishing baryon number have $B=\pm 1$.

\begin{figure*}[tb]
   \includegraphics[width=0.49\linewidth]{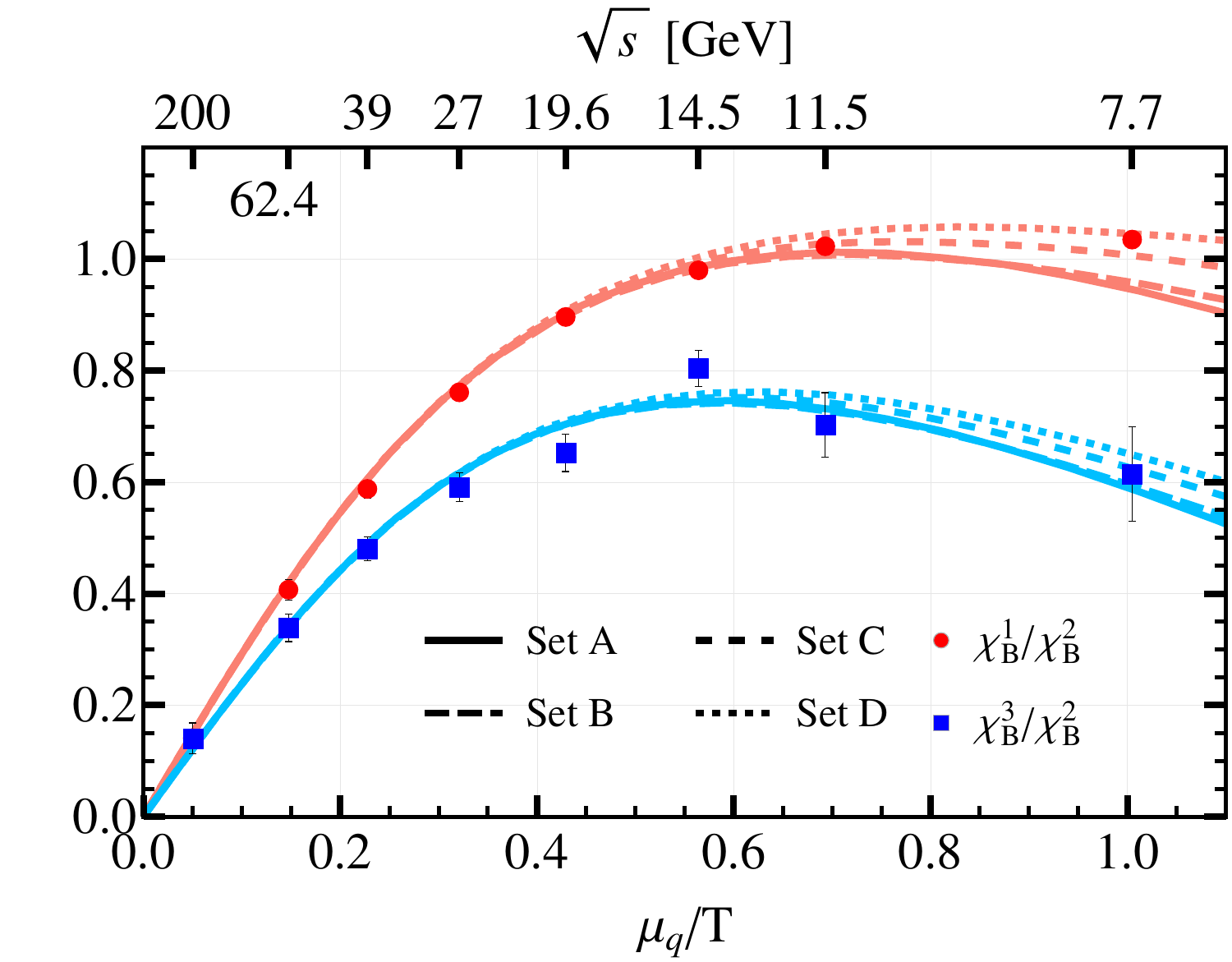}
	\includegraphics[width=0.49\linewidth]{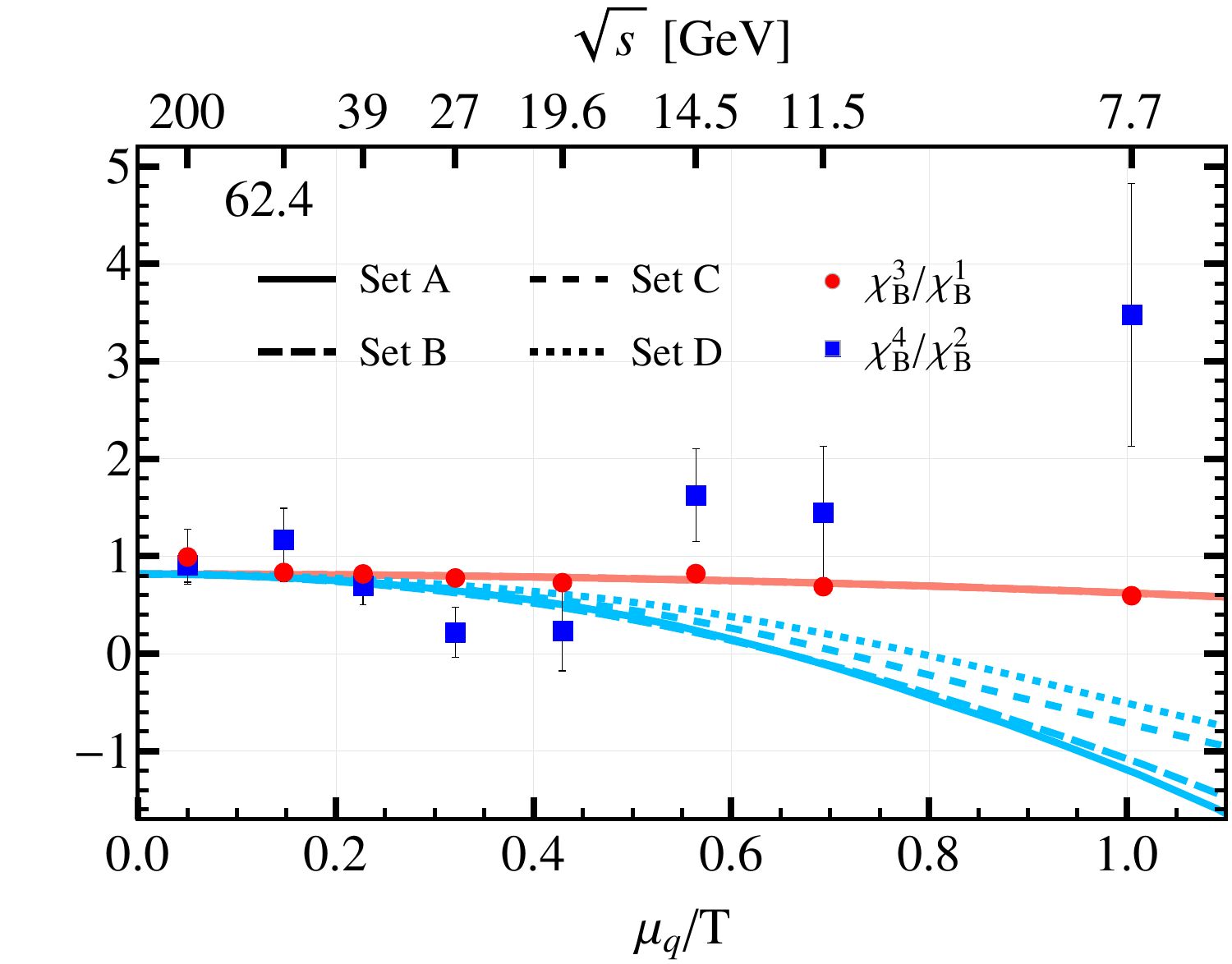}
   \caption{
   Ratios of cumulants of net-baryon-number fluctuations  in the PQM model  along the freeze-out line, obtained by fitting $\chi_B^3/\chi_B^1$ to the STAR data.  The four sets of model parameters used and the preliminary STAR data shown, are the same as in Fig. \ref{fig:five}.
 \label{fig:six}}
\end{figure*}

A comparison of results obtained in the PQM model with data requires a correspondence between the collision energy $\sqrt s$ and the thermal parameters $(\mu_q, T)$. Here we employ the phenomenological relation,  obtained by analysing the freeze-out conditions in terms of the hadron-resonance-gas model (HRG) \cite{BraunMunzinger:2003zd,Andronic:2005yp,Andronic:2008gu,Cleymans:2005xv}. We then  use the  resulting dependence of $\mu_B$ and $T$ on $\sqrt s$ to assign a value for the ratio $(\mu_q/T)$ to each of the STAR beam energies. We note that, for $\mu_q/T<1$,  the phenomenological   freeze-out line coincides within errors with the crossover  phase boundary obtained in   lattice QCD  \cite{Borsanyi:2010bp,Andronic:2008gu}. This  motivates a comparison of model results  on net-baryon-number  fluctuations near  the phase boundary with data. Such an analysis was first done using LQCD results  in Ref.~\cite{Bazavov:2012vg}.

In Fig.~\ref{fig:five}, we show the STAR data on net-proton-number  susceptibility ratios and the corresponding PQM model results on net-baryon-number fluctuations computed along the phase boundary. The model results for the ratios $\chi_B^{1,2}$, $\chi_B^{3,1}$ and $\chi_B^{3,2}$ are in qualitative agreement with the data in the whole energy range. For the kurtosis ratio, $\chi_B^{4,2}$, this is the case  also  up to the SPS energy, i.e., for $\sqrt s \geq 20$ GeV. However,  for $\mu_q/T>0.5$,  the data on the kurtosis ratio exhibits a qualitatively different dependence on $\mu_q/T$  than  expected for the critical behavior of $\chi_B^{4,2}$, as the CEP is approached along the phase boundary.

As noted above,  the ratio $\chi_B^{1,2}$ is, on the phase boundary, approximately given by $\chi_B^{1,2}\simeq \tanh (3\mu_q/T)$ up to $\mu_q/T\simeq 1$, as seen in Fig. \ref{fig:five}. This form of the ratio of the lowest order cumulants is also obtained in the HRG model and is consistent with  LQCD~\cite{Karsch:2010ck,Allton:2005gk}.
Fig. \ref{fig:five} reveals that  the data are  consistent with  this form as well.  This fact clearly supports the use of the   $\sqrt s$ dependence of $\mu_q/T$ obtained from  the HRG model analysis of multiplicities in the comparison of the PQM  model results with data.

In order to assess  the sensitivity of the fluctuation observables  to model parameters,  we show  in Fig. \ref{fig:five} ratios of cumulants   for four  sets of model parameters. The different sets are obtained by varying the sigma meson mass and the form of the Polyakov-loop potential.  The  parameter sets are described in  Appendix \ref{sec:initc}.
Fig. \ref{fig:five} shows that, although some quantitative  differences can be identified at larger $\mu_q/T$, the skewness, kurtosis  and $\chi_B^{1,2}$ ratios along the phase boundary are  qualitatively similar for the different sets of model parameters.

In the comparison of model predictions with data in Fig.  \ref{fig:five}, we assume,  that the freeze-out of the net-baryon-number  fluctuations,  tracks the chiral phase boundary. Clearly, this simple assumption provides a qualitative understanding of the data. In order to obtain a more quantitative description, we follow
Refs.~\cite{Bazavov:2012vg} and  \cite{Almasi:2016hhx,Fu:2015amv}, and  determine the  freeze-out conditions by fitting the data on the $\chi_B^{3,1}$ ratio, using the $(\sqrt s)$-dependence  of $\mu_q/T$ obtained from  the fit of the  HRG model to particle multiplicities~\cite{Andronic:2005yp,Andronic:2008gu,Cleymans:2005xv}.

\begin{figure*}[tb]
   \includegraphics[width=0.49\linewidth]{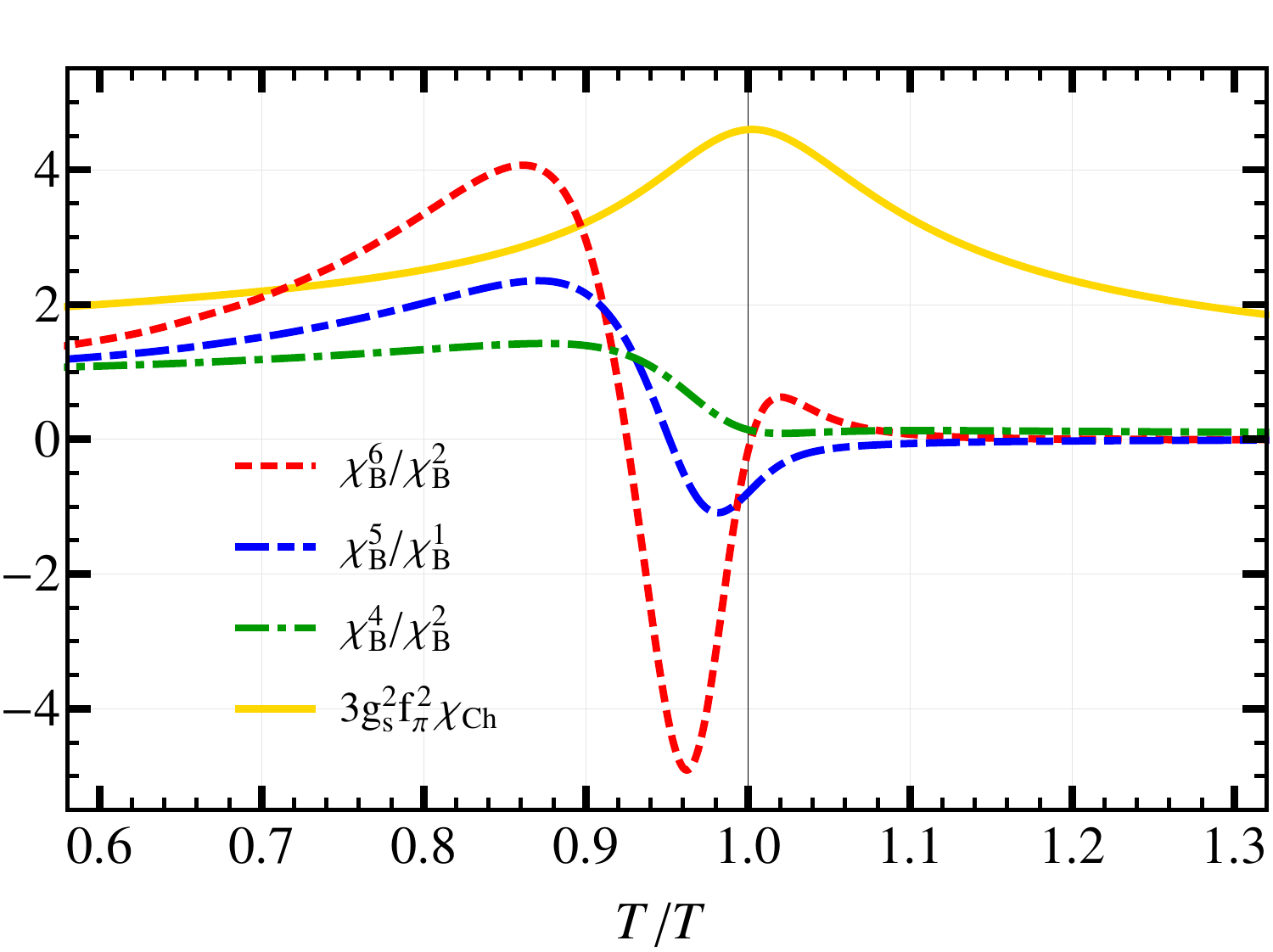}
	\includegraphics[width=0.49\linewidth]{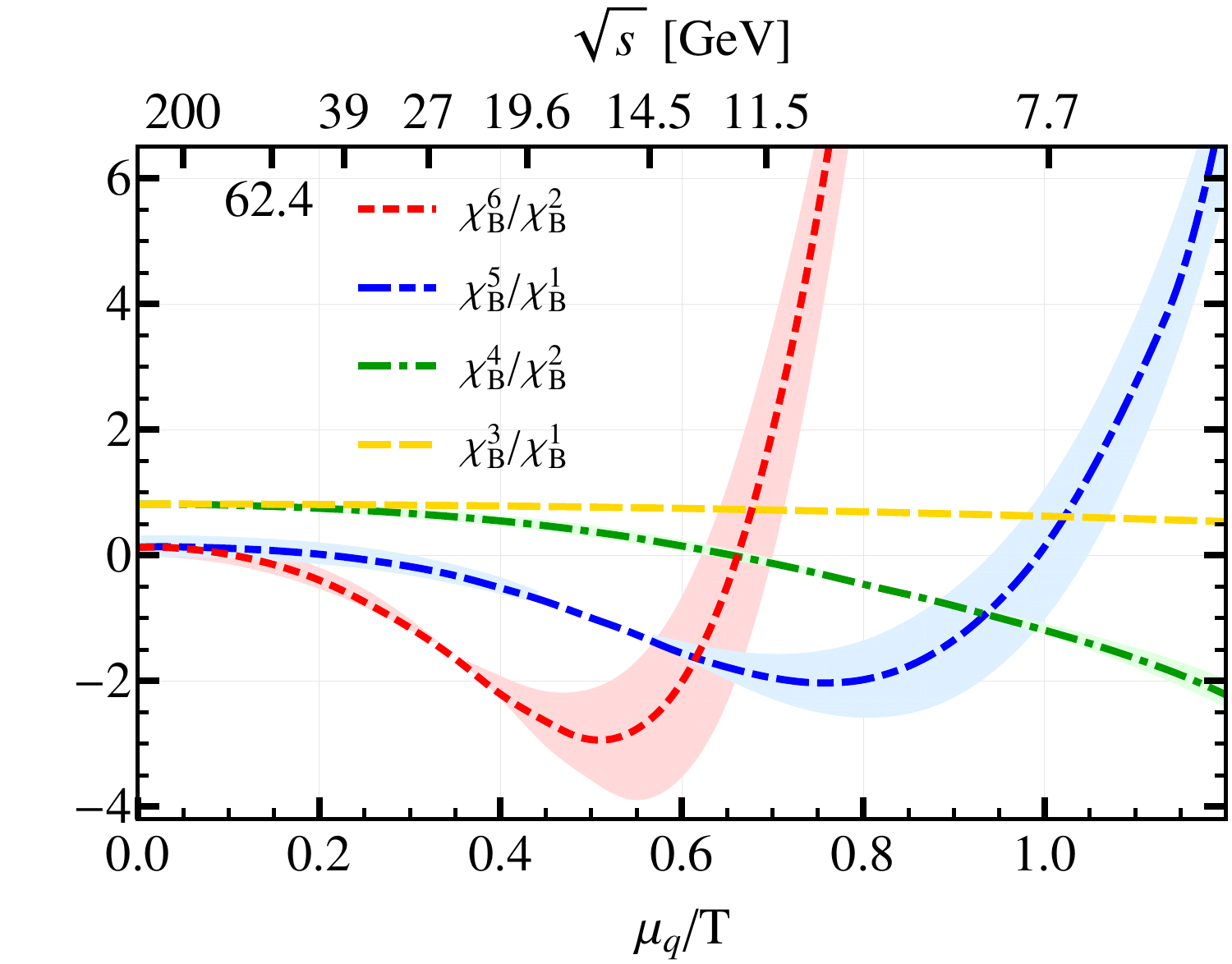}
\caption{Left-hand figure: The temperature dependence of ratios of net-baryon-number   cumulants and the chiral susceptibility $\chi_{ch}$, all   computed in the PQM model at  $\mu_q=0.1$ GeV.
 The vertical line indicates the location  of the chiral crossover temperature.
 Right-hand figure:
The ratios $\chi_B^4/\chi_B^2$, $\chi_B^5/\chi_B^1$ and $\chi_B^6/\chi_B^2$ along the freeze-out line obtained by fitting $\chi_B^3/\chi_B^1$ (also shown) to data. These results were computed using the model parameters of set A  (see Appendix \ref{sec:initc}). The bands about $\chi_B^5/\chi_B^1$ and $\chi_B^6/\chi_B^2$ reflect the experimental error in $\chi_B^3/\chi_B^1$, which leads to an uncertainty in the freeze-out temperature.
  \label{fig:seven}}
\end{figure*}

In  Fig.~\ref{fig:six} we show the fluctuation ratios along the freeze-out line, which is fixed through the skewness  data. The model results are obtained for the four sets of initial conditions. Fig.~\ref{fig:six} clearly shows that, along the freeze-out line, the spread of all fluctuations ratios considered  for the various parameter sets is much weaker than that observed in Fig. \ref{fig:five} along the phase boundary.  This indicates, that moderate changes of the sigma mass and modifications of the form of the Polyakov loop potential may lead to a shift in the temperature scale but essentially  with no change of the relative structure of the cumulant ratios.

The results presented in Fig. \ref{fig:six} clearly show that the model provides a very good description of the data on $\chi_B^{1,2}$ and   $\chi_B^{3,2}$. Also the  kurtosis data,  obtained at higher  collision energies,  are consistent with model results.  However,  at $\sqrt s<  20$ GeV they exhibit a different  trend, with the data increasing rapidly at lower energies, while the model result keeps decreasing. We conclude that  an   increase of  $\chi_B^{4,2}$  ratio beyond unity,  observed in the STAR data at $\sqrt s< 20$ GeV, is not expected in equilibrium on the chiral critical line  nor on the freeze-out line.

As noted above, the ratio $\chi_B^{6,2}$ is particularly interesting for identifying criticality governed by the $O(4)$ universality class.  This is seen in Fig. \ref{fig:one} at $\mu_q=0$,  where $\chi_B^{6,2}$ is negative at $T=T_c$. At $\mu_q>0$, the influence of criticality is even  more pronounced, as shown in Fig.~\ref{fig:seven}. There,    $\chi_B^{6,2}$ exhibits a highly nonmonotonic structure near the phase boundary.

In the left panel of Fig.~\ref{fig:seven}  we show the temperature dependence of the ratio $\chi_B^{6,2}$ near the phase boundary for  $\mu_q=0.1$ GeV. This ratio exhibits a very strong variation near the transition temperature.  It is deeply  negative below $T_c$ and develops a maximum just above $T_c$. The value of $\chi_B^{6,2}$ is thus very sensitive to the freeze-out temperature.

On the right of Fig.~\ref{fig:seven}  we show the ratios  $\chi_B^4/\chi_B^2$,  $\chi_B^5/\chi_B^1$ and  $\chi_B^6/\chi_B^2$ computed, using the model parameters of set A,  along  the freeze-out line which is determined by fitting $\chi_B^3/\chi_B^1$ to data. At vanishing $\mu_q$,  $\chi_B^4/\chi_B^2=\chi_B^3/\chi_B^1$ and $\chi_B^6/\chi_B^2=\chi_B^5/\chi_B^1$, as a consequence of Eq.~\eqref{eq:odd-even}, while at larger $\mu_q/T$, these ratios separate.

We note at this point,  that the equality of the ratios $\chi_B^4/\chi_B^2$ and $\chi_B^3/\chi_B^1$ in the STAR data at the highest energy is a strong indication that the fluctuations probed by these cumulants are in thermal equilibrium. It is very unlikely that a system not in equilibrium would yield ratios of cumulants that satisfy Eq.~\eqref{eq:odd-even}. Note,  that at $\mu_q=0$, the critical $O(4)$ fluctuations yield divergent contributions only to $\chi^6_B$ and higher cumulants~\cite{Friman:2011pf}. Thus, the fluctuations probed by $\chi_B^4/\chi_B^2$ and $\chi_B^3/\chi_B^1$ are not critical. However, a measurement of the ratios involving the fifth and sixth order cumulants would probe whether the $O(4)$ critical fluctuations are in equilibrium or not.  Obviously, this test of equilibration is meaningful only at small $\mu_q/T$, i.e., only at the highest energies.

At moderate values of $\mu_q/T$, the $\chi_B^6/\chi_B^2$ ratio is negative and deviates clearly from $\chi_B^5/\chi_B^1$. At still lower energies, it  exhibits a strong increase towards the CEP, where it diverges. Similarly, $\chi_B^5/\chi_B^1$ decreases at moderate $\mu_q/T$, and increases strongly as the CEP is approached.  These results indicate that in heavy ion collisions $\chi_B^6/\chi_B^2$ and $\chi_B^5/\chi_B^1$ will exhibit strong nonmonotonic dependencies on $\sqrt s$.

Recently,  first results on $\chi_B^6/\chi_B^2$  in Au-Au collisions  at $\sqrt{s}=200\;\mathrm{GeV}$ where reported by the STAR Collaboration for  several centralities~\cite{Eshatalk}. The data show a strong suppression of fluctuations compared to the kurtosis ratio. In mid-central and the most central  collisions, the $\chi_B^6/\chi_B^2$ fluctuation ratio  is negative, albeit with still very large statistical  uncertainties. In fact, given the large errors for the most central collisions, the preliminary data is consistent with a vanishing $\chi_B^6$. A value close to zero is consistent  with the model results, shown in the right panel of Fig.~\ref{fig:seven}.

The  comparison of model results on ratios of net-baryon-number susceptibilities with the STAR data in Figs.~\ref{fig:five}, \ref{fig:six}, and \ref{fig:seven}, shows that the data, with the exception of kurtosis at low energies, follow  general trends expected due to critical chiral dynamics. We note,  that the ratios of net-baryon-number susceptibilities  near the phase boundary involving net-baryon number cumulants $\chi_B^n$ with $n\geq 3$  are  controlled mainly by the  scaling functions in the $O(4)$ and $Z(2)$ universality classes, respectively. This observation indicates, that by measuring fluctuations of conserved charges in heavy-ion collisions,  we are indeed probing the QCD phase boundary, and thus accumulating evidence for chiral symmetry restoration.

However, as discussed above, there are several uncertainties and assumptions which must be thoroughly understood before the QCD phase boundary can be pinned down with confidence. Possible contributions to fluctuation observables from effects not related with critical phenomena, like e.g. baryon-number conservation~\cite{Bzdak:2012an}  and volume fluctuations~\cite{Skokov:2012ds,Braun-Munzinger:2016yjz, Palhares:2009tf,Fraga:2011hi,Palhares:2012zz,Hippert:2015rwa} are being explored.
We note, however,  that volume fluctuations do not modify the double ratios proposed as tests of equilibration at $\mu_q\to 0$.
We also mention the  rather  strong sensitivity  of higher order net-proton-number cummulants on the transverse momentum range imposed in the analysis of the STAR data.  Nevertheless, it is intriguing,  that the dynamics of the model, provides a good description of the STAR data (except for $\chi_B^4$ at the lowest energies), without all these effects of non-critical origin. It remains an important task to assess the effect of theses additional sources of fluctuations in the whole energy range probed by the expriments.


\section{Summary and Conclussions\label{sec:summary}}

We have studied  the influence of the chiral phase transition  on fluctuation observables in   strongly interacting medium at finite temperature and net baryon number density. We focused on the properties of net-baryon-number fluctuations which are quantified by the $n^{\mathrm{th}}$-order  susceptibilities, $\chi_B^n$. The cumulants $\chi_B^n$  are directly influenced by the chiral phase transition due to the coupling of the quarks to the scalar sigma field. Furthermore, since the $\chi_B^n$ are accessible experimentally, they are ideal observables to identify the phase boundary  and the critical structures in the QCD phase diagram.

The dynamics was modelled with the Polyakov loop extended Quark-Meson (PQM) model. To correctly account for the critical behavior at the chiral symmetry restoration transition in the $O(4)$ and $Z(2)$ universality classes,  we employed  the Functional Renormalization Group (FRG).  We formulated the FRG equations in the presence  of repulsive interactions, and  derived the flow equations for  derivatives of the thermodynamic pressure.

The main point of our studies was  to identify the relations between  $\chi_B^n$ susceptibilities  in different $(T,\mu_B)$ - regions of the phase diagram.
To reduce   the  influence of the  non-critical characteristics of the model, like e.g. the  mass spectrum or the  kinematical cuts on particle momentum distributions,  we have computed ratios of susceptibilities. Here,  of particular interest,  are  ratios of the first and second order cumulants, $\chi_B^1/\chi_B^2$,  the skewness, $\chi_B^3/\chi_B^1$, the   kurtosis, $\chi_B^4/\chi_B^2$, and the sixth-order cumulant $\chi_B^6/\chi_B^2$. The higher-order ($n\geq 3$) cumulants are  probes of chiral  criticality  and can be reconstructed  experimentally in nucleus-nucleus collisions.

We have calculated, for the first time,  the  contour plots in the $(T,\mu_B)$-plane of  the above ratios in the PQM model within  FRG approach. We have quantified the systematic relations along  the phase boundary and the phenomenological freeze-out line,  extracted by fitting the skewness data. The influence of the repulsive interactions and the position of the critical endpoint on different $\chi_B^n$ ratios, was also explored for the first time.

The model results were confronted with ratios of cumulants of net proton number measured by the STAR
Collaboration in nucleus-nucleus collisions at energies ranging from $\sqrt s = 7.7$ GeV up to  200 GeV.

Considering that  the relations between different  susceptibilities  along the phase boundary are induced in part by the universal scaling functions common to QCD and  PQM models,  and the phenomenological observation that freeze-out in heavy-ion collisions appears   near the  QCD  phase boundary,  we have compared the experimental data and model results. The main objective was to verify if the systematics between net-baryon-number susceptibilities observed  in the model calculations along the phase boundary are also reflected in the experimental data. To compare  model predictions and  data we have assumed that   the  $\sqrt s$-dependence of $\mu_q/T$,  follows  the phenomenological freeze-out conditions in nucleu-nucleus collisions. The validity of this assumption is supported by the observation that, for $\mu_q/T <1$, the ratio of the first  and second  order cumulants  is well approximated   by   $\chi_B^1/\chi_B^2= \tanh(3\mu_q/T)$,  in the model, in LQCD and in the data, using the adopted relation between beam energy and $\mu_q/T$.

We have shown, that the STAR data for the ratios composed of  $\chi_B^n$ with $n=1,2,3$ follow   generic  expectations, and thus that they are consistent with criticality at the chiral  phase boundary. The quantitative agreement between   the calculated cumulant ratios and  STAR  data is  improved  when the  freeze-out line is determined by fitting $\chi_B^3/\chi_B^1$ to  data. The data on the kurtosis ratio are  also  consistent with model systematics at energies ${\sqrt{s} \geq 20\;\mathrm{GeV}}$. However, the strong enhancement of $\kappa\sigma^2>1$, observed in nucleus-nucleus collisions at lower energies, is not reproduced by the model along the path  in the $(\mu_q/T)$-  plane,  where the ratios of lower order cumulants, $\chi_B^n$ with $n=1,2,3$, are well described. We have shown,  that this conclusion is   not affected by  the initial conditions nor by the  parametrization of the Polyakov-loop potential.

Clearly, the comparison of model results  and data is biased  by the various assumptions and uncertainties, discussed  in Sec.~\ref{sec:phaseboundary}.
However, in spite of all uncertainties,  the ratios of net-proton-number susceptibilities  obtained by
STAR   are, as shown in this study,  largely   consistent with the  systematics expected near  the chiral  phase boundary.

It is interesting to note that a recent analysis of the STAR data on skewness and kurtosis ratios,  at $\sqrt s \geq  19. 6$ GeV, using LQCD results for the  Taylor series in $\mu_q/T$,  shows  that the data exhibits all features expected in QCD near the phase boundary \cite{Karsch:2016yzt}. This result support our findings in PQM model  that  near  chiral phase boundary, ratios of net-baryon-number cumulants,    capture   properties,  which are common   to QCD and the PQM model.

\section*{Acknowledgments}
 We acknowledge stimulating discussions with Peter Braun-Munzinger, Frithjof Karsch and  Nu Xu. G.A. is also grateful to Vladimir Skokov for  discussions on   vector interactions. The work of B.F. and K.R. was partly supported
by the Extreme Matter Institute EMMI.
K. R. also  acknowledges  partial  supports of the Polish National Science Center (NCN) under Maestro grant DEC-2013/10/A/ST2/00106.
G. A. acknowledges the support of  the Hessian LOEWE initiative
through the Helmholtz International Center for FAIR (HIC for FAIR).

\appendix

\begin{widetext}
\section{Flow equations  for   derivatives of the thermodynamic potential  \label{sec:flows}}
In the following,   we introduce the flow equations that are used  to  calculate  $\chi_B^1$ and $\chi_B^2$ cumulants of net-baryon-number fluctuations. We start from the flow equation   (\ref{eq:floweq}) for the grand canonical potential,
\begin{equation}
	\frac{d}{dk}\Omega_k=3 F_{b}\left( m_{\pi}^2 \right) + F_{b}\left( m_{\sigma}^2 \right) + F_{f}\left(\mu,\Phi,\bar{\Phi}\right),\label{flow}
\end{equation}
where the boson and fermion contribution is introduced,  as
\begin{align}
	F_{b}(m^2) &= \frac{k^4}{12\pi^2} \frac{\coth\left( \frac{\sqrt{k^2+m^2}}{2T} \right)}{\sqrt{k^2+m^2}}, \\
	F_{f}(\mu,\Phi,\bar{\Phi}) &= -\frac{k^4 N_c N_f}{3\pi^2 E_q}\big( 1-N(T,\mu;\sigma,\Phi,\bar{\Phi}) -N(T,-\mu;\sigma,\bar{\Phi},\Phi) \big)
\end{align}
with the pion and sigma masses obtained from
\begin{equation}
	m_{\pi}^2 = \frac{1}{\sigma}\frac{\partial \Omega_k}{\partial \sigma},\quad\quad\quad
	m_{\sigma}^2 = \frac{\partial^2 \Omega_k}{\partial \sigma^2}.
\end{equation}
The flow equations for  derivatives, introduced in Eq.   (\ref{eq:flows}), are derived for each gridpoint from Eq. (\ref{flow}), as
\begin{align}
\begin{split}
	\frac{d}{dk} \frac{\partial \Omega_k}{\partial \lambda_i} &=
	\frac{\partial}{\partial \lambda_i} \frac{d \Omega_k}{dk}  =
	3 F_{b}'\left(m_{\pi}^2\right)\frac{1}{\sigma}\frac{\partial^2\Omega_k}{\partial\lambda_i\partial\sigma}+
	 F_{b}'\left(m_{\sigma}^2\right)\frac{\partial^3\Omega_k}{\partial\lambda_i\partial^2\sigma}
	+\frac{\partial}{\partial \lambda_i}F_{f}(\mu,\Phi,\bar{\Phi}), \\
	\frac{d}{dk} \frac{\partial^2 \Omega_k}{\partial \lambda_i \partial \lambda_j} &=
	\frac{\partial^2}{\partial \lambda_i \partial \lambda_j} \frac{d \Omega_k}{dk}  =
	3 F_{b}''\left(m_{\pi}^2\right)\frac{1}{\sigma^2}\frac{\partial^2\Omega_k}{\partial\lambda_i\partial\sigma}\frac{\partial^2\Omega_k}{\partial\lambda_j\partial\sigma}
	+3 F_{b}'\left(m_{\pi}^2\right)\frac{1}{\sigma}\frac{\partial^3\Omega_k}{\partial\lambda_i \partial\lambda_j \partial\sigma}
	\\ &+
	 F_{b}''\left(m_{\sigma}^2\right)\frac{\partial^3\Omega_k}{\partial\lambda_i\partial^2\sigma}
	\frac{\partial^3\Omega_k}{\partial\lambda_j\partial^2\sigma}
	 +F_{b}'\left(m_{\sigma}^2\right)\frac{\partial^4\Omega_k}{\partial\lambda_i\partial\lambda_j\partial^2\sigma}
	+\frac{\partial^2}{\partial \lambda_i \partial \lambda_j}F_{f}(\mu,\Phi,\bar{\Phi}),
\end{split}
\end{align}
where $\lambda_i$ and $\lambda_j$ stand for   one of the $\lbrace \mu,\Phi,\bar{\Phi}\rbrace$ parameters, and the prime denotes derivative with respect to $m^2$.\newpage
\end{widetext}

\section{Polyakov-loop potential and initial conditions \label{sec:initc}}

\begin{table}
\begin{tabular}{ |c| c| c| c| c| }
\hline
  $a_1$ & $a_2$ & $a_3$ & $a_4$ & $a_5$ \\ \hline
  -44.14 & 151.4 & -90.0677 & 2.77173 & 3.56403 \\ \hline
  $b_1$ & $b_2$ & $b_3$ & $b_4$ &   \\ \hline
  -0.32665 & -82.9823 & 3.0 & 5.85559 & \\ \hline
  $c_1$ & $c_2$ & $c_3$ & $c_4$ & $c_5$ \\ \hline
  -50.7961 & 114.038 & -89.4596 & 3.08718 & 6.72812 \\ \hline
  $d_1$ & $d_2$ & $d_3$ & $d_4$ & $d_5$ \\ \hline
  27.0885 & -56.0859 & 71.2225 & 2.9715 & 6.61433 \\ \hline
\end{tabular}
\caption{Parameters of the Polyakov-loop potential  introduced in Eqs. (\ref{our}) and (\ref{par}). \label{tab:polyconsts}}
\end{table}

\begin{table}
\begin{tabular}{ |c| c| c| c| c| }
\hline
  SET & Initial conditions & Polyakov-loop potential  \\ \hline
   A &  (\ref{B8}) & (\ref{our}) \\ \hline
   B & (\ref{B8}) & (\ref{pot1})\\ \hline
   C & (\ref{B9}) &  (\ref{our}) \\ \hline
   D &  (\ref{B9}) &  (\ref{pot1}) \\ \hline
\end{tabular}
\caption{Different combinations of the initial conditions and the Polyakov-loop potentials,   used in the PQM model.\label{tab:runs}}
\end{table}

To verify  possible   model dependence of our results,  we have considered   different parameterizations of the Polyakov-loop potential.

We apply   the polynomial Polyakov-loop potential \cite{Ratti:2005jh},    as
\begin{equation}
	\frac{\mathcal{U}(\Phi,\bar{\Phi};T)}{T^4} = -\frac{b_2(T)}{2}\left( \Phi\bar{\Phi} \right) - \frac{b_3}{6} \left( \Phi^3 + \bar{\Phi}^3 \right)
	+\frac{b_4}{4}\left(\Phi\bar{\Phi}\right)^2,\label{pot1}
\end{equation}
where  $b_3=0.75$, $b_4=7.5$,   and
\begin{equation}
	b_2(T)=a_0 + a_1 \left(\frac{T_0}{T}\right)
	+ a_2 \left(\frac{T_0}{T}\right)^2
	+ a_3 \left(\frac{T_0}{T}\right)^3,
\end{equation}
with $a_0=6.75$, $a_1=-1.95$, $a_2=2.625$, $a_3=-7.44$, and  $T_0=270\, \mathrm{MeV}$.

 The parameter $T_0$ corresponds to the value of the critical temperature at the  first order deconfinement phase transition in  a pure Yang-Mills theory. Effects of unquenching can be taken into account by changing  $T_0$. In the  case of two quark flavors, we use  $T_0=208\, \textrm{MeV}$ from Ref.  \cite{Herbst:2010rf}.

The second potential applied in our calculations, accounts for the color group structure,  and reproduces  lattice results on the equation of state and fluctuations of the Polyakov loops,  in a pure SU(3) gauge theory.   The potential is parameterized \cite{Lo:2013hla},  as
\begin{align}
	\frac{\mathcal{U}(\Phi,\bar{\Phi};T)}{T^4} = &-\frac12 a(T)\left( \Phi\bar{\Phi} \right)
	+ b(T) \log M_H(\Phi,\bar{\Phi})
	\nonumber \\
	 &+ \frac12 c(T) \left( \Phi^3 + \bar{\Phi}^3 \right) +d(T)\left(\Phi\bar{\Phi}\right)^2,\label{our}
\end{align}
where
\begin{equation}
	M_H(\Phi,\bar{\Phi})=\left(1-6\Phi\bar{\Phi}
	+ 4\left( \Phi^3 + \bar{\Phi}^3 \right) -3\left(\Phi\bar{\Phi}\right)^2 \right).
\end{equation}
The temperature dependent coefficients are given by
\begin{align}
\begin{split}
	a(T)&=\frac{a_1 t^2 + a_2 t +a_3}{t^2+a_4 t + a_5},\quad b(T)= b_1 t^{-b_4}\left(1-e^{b_2/t^{b_3}}\right),\\
	c(T)&=\frac{c_1 t^2 + c_2 t +c_3}{t^2+c_4 t + c_5},\quad d(T)=\frac{d_1 t^2 + d_2 t +d_3}{t^2+d_4 t + d_5},\label{par}
\end{split}
\end{align}
where $t=T/T_0$. The numerical values of the constants  in Eq. (\ref{par}),  are summarized in Tab.~\ref{tab:polyconsts}. We set ${T_0=270\; \textrm{MeV}}$ for this potential. Lower values of $T_0$ would result in distinct position of the  peaks of  the Polyakov loop and chiral susceptibilities, which is in contrast with LQCD results.

The initial conditions for the flow equations are determined by the vacuum conditions:
\begin{align}
	\langle&\sigma\rangle_{T=0,\mu=0} = f_{\pi}=93 \; \mathrm{MeV}, \nonumber\\
	&m_{\pi}=138 \; \mathrm{MeV}, \quad\quad\quad
	m_q=300 \; \mathrm{MeV}.
\end{align}
The Yukawa coupling $g_s$ and the external field are then determined from
\begin{equation}
	g_s=m_q/f_{\pi},\quad\quad\quad H=m_{\pi}^2 f_{\pi}.
\end{equation}
However, the values of $\lambda$ and $m^2$,  in the initial meson potentials, are still undefined. They can be fixed by specifying   the cutoff scale $\Lambda$, where the flow is started, and the mass of the sigma meson,  $m_{\sigma}$.  We apply,    two  distinct values of the initial cutoff scales,  and the sigma  masses, as
\begin{align}
	\Lambda&=700\; \textrm{MeV},\label{B8} \\m_{\sigma}&\approx 410\; \textrm{MeV},\quad
	  \lambda = -1, \quad m^2 = 2.9\cdot 10^5 \; \mathrm{MeV}^2, \nonumber\\
	\Lambda&=950\; \textrm{MeV},\label{B9}\\ m_{\sigma}&\approx 500\; \textrm{MeV},\quad
	  \lambda = 1.3, \quad m^2 = 5.668\cdot 10^5 \,  \mathrm{MeV}^2.\nonumber
\end{align}
where the  parameters in    Eq. (\ref{B9}),  were introduced in  Ref. \cite{Herbst:2013ail}.

From the two parametrization of the   Polyakov-loop potentials,  Eqs. (\ref{pot1}) and (\ref{our}),  and the two chiral initial conditions  defined in Eqs. (\ref{B8}) and (\ref{B9}), we define in  Table~\ref{tab:runs}, the four different  parameter sets that are used in our calculations to study the influence of the model assumptions on  the final results.    If the parameter set is not explicitly specified, then results refer  to set A.

\bibliography{refs}

\end{document}